\documentclass{aa}  

\usepackage{graphicx}
\usepackage{txfonts}
\usepackage{subfigure}
\usepackage{color}
\usepackage[none]{hyphenat}

\def\thetad{\dot{\theta}}
\def\norb{n}
\def\res{p}
\def\gam{\gamma}
\def\gamd{\dot{\gam}}
\def\C{C}
\def\Ttriax{T_\mathrm{tri}}
\def\Ttide{T_\mathrm{tid}}
\def\Todd{T_\mathrm{tid}^\mathrm{odd}}
\def\Teven{T_\mathrm{tid}^\mathrm{even}}
\def\Hpe{H_{p,e}}
\def\Pcapt{P_\mathrm{cap}}
\def\width{\Delta}
\def\Tm{\tau_\mathrm{M}}
\def\d{\mathrm{d}}

\newcommand{\balign}[1]{
\begin{align}
#1
\end{align}}

\newcommand{\eq}[1]{Eq.\,(\ref{#1})}

\newcommand{\fig}[1]{Fig.\,\ref{#1}}

\begin{document}

   \title{The habitability of Proxima Centauri b}
   \subtitle{I. Irradiation, rotation and volatile inventory from formation to the present}

   \author{Ignasi Ribas\inst{1}
          \and
          Emeline Bolmont\inst{2}
          \and
          Franck Selsis\inst{3}
          \and
          Ansgar Reiners\inst{4}
           \and
          J\'er\'emy Leconte\inst{3}
          \and
          Sean N. Raymond\inst{3}
          \and
          Scott G. Engle\inst{5}
          \and
          Edward F. Guinan\inst{5}
          \and
          Julien Morin\inst{6}
          \and 
          Martin Turbet \inst{7}
          \and
          Fran\c cois Forget \inst{7}
          \and
          Guillem Anglada-Escud\'e\inst{8}
          }

   \institute{Institut de Ci\`encies de l'Espai (IEEC-CSIC), C/Can Magrans, 
s/n, Campus UAB, 08193 Bellaterra, Spain\\
              \email{iribas@ice.cat}
         \and
NaXys, Department of Mathematics, University of Namur, 8 Rempart de la Vierge,
5000 Namur, Belgium
         \and
Laboratoire d'astrophysique de Bordeaux, Univ. Bordeaux, CNRS, B18N, all\'ee Geoffroy 
Saint-Hilaire, 33615 Pessac, France
         \and
Institut f\"ur Astrophysik, Friedrich-Hund-Platz 1, 37077 G\"ottingen, Germany 
         \and
Department of Astrophysics and Planetary Science, Villanova University,
Villanova, PA 19085 USA
         \and
LUPM, Universit\'e de Montpellier, CNRS, Place E. Bataillon, 34095 
Montpellier, France
         \and
Laboratoire de M\'et\'eorologie Dynamique, IPSL, Sorbonne Universit\'es, UPMC Univ Paris 06, 
CNRS, 4 place Jussieu, 75005 Paris, France
         \and
School of Physics and Astronomy, Queen Mary University of London, 327 Mile End 
Rd, London E1 4NS, UK
             }

   \date{Received; accepted}

 
  \abstract
{Proxima b is a planet with a minimum mass of $1.3$~M$_\oplus$ orbiting within the 
habitable zone (HZ) of Proxima Centauri, a very low-mass, active star and the Sun's closest 
neighbor. Here we investigate a number of factors related to the potential habitability 
of Proxima b and its ability to maintain liquid water on its surface. We set the stage by 
estimating the current high-energy irradiance of the planet and show that the planet 
currently receives 30 times more EUV radiation than Earth and 250 times more X-rays. We compute 
the time evolution of the star's spectrum, which is essential for modeling the 
flux received over Proxima b's lifetime. We also show that Proxima b's obliquity is likely 
null and its spin is either synchronous or in a 3:2 spin-orbit resonance, depending on the 
planet's eccentricity and level of triaxiality. Next we consider the evolution of Proxima 
b's water inventory. We use our spectral energy distribution to compute the hydrogen loss 
from the planet with an improved energy-limited escape formalism. Despite the high level 
of stellar activity we find that Proxima b is likely to have lost less than an Earth 
ocean's worth of hydrogen (EO$_H$) before it reached the HZ 100--200 Myr after its formation.  
The largest uncertainty in our work is the initial water budget, which is not constrained 
by planet formation models. We conclude that Proxima b is a viable candidate habitable 
planet.}  


   \maketitle
%

\section{Introduction}

The discovery and characterization of Earth-like planets is among the most exciting 
challenges in science today.  A plethora of rocky planets have been discovered in recent 
years by both space-based missions such as {\it Kepler} \citep{borucki10,batalha13} and 
by ground-based radial velocity monitoring \citep{mayor11}. \citet{Anglada16} have 
announced the discovery of Proxima b, a planet with a minimum mass of 1.3~M$_{\oplus}$ 
orbiting Proxima Centauri, the closest star to the Sun. Table \ref{Char_prox} shows 
the characteristics of Proxima and its discovered planet.

Here -- as well as in a companion paper \citep[][hereafter Paper~II]{Turbet16} -- we 
address a number of factors related to the potential habitability of Proxima b.  

Defining planet habitability is not straightforward. In the context of the search for 
signs of life on exoplanets, the presence of stable liquid water on a planet's surface 
represents an important specific case of habitability. There are strong thermodynamic 
arguments to consider that the detection of a biosphere that is confined into a planetary 
interior with no access to stellar light will require in-situ exploration and may not be 
achieved by remote observations only \citep{Rosing2005}. Surface habitability requires 
water but also an incoming stellar flux low enough to allow part of the water to be in 
the liquid phase but sufficient to maintain the planetary surface (at least locally) 
above 273~K. These two limits in stellar flux determine the edges of the Habitable Zone 
(HZ) as defined by \citet{Kasting1993}. Proxima b orbits its star at a distance 
that falls well within its HZ limits, with a radiative input of 65--70\% of the Earth's 
value ($S_\oplus$) based on the measured orbital period, and on estimates of the 
stellar mass \citep{Delfosse2000} and bolometric luminosity 
\citep{Demory2009,Boyajian2012}. The inner and outer limits of a conservative HZ are 
indeed estimated at 0.9 and 0.2~$S_\oplus$, respectively \citep{Kopparapu2013}. For 
a synchronized planet, the inner edge could be as close as 1.5~S$_\oplus$ 
\citep{Yang2013,Kopparapu2016}.

Although Proxima b's insolation is similar to Earth's, the context of its habitability 
is very different. Proxima is a very low mass star, just 12\% as massive as the Sun. 
Proxima's luminosity changed considerably during its early evolution, after Proxima b had 
already formed. As a consequence, and in contrast with the evolution of the Solar System, 
the HZ of Proxima swept inward as the star aged. Proxima b spent a significant amount of 
time interior to the HZ before its inner edge caught up with the planet's orbit 
\citep[e.g.,][]{RamirezKaltenegger2014}. This phase of strong irradiation has the 
potential to induce water loss, with the potential for Proxima b entering the HZ as 
dry as present-day Venus.  We return to this question in Sect. \ref{co_evol}. Rotation 
represents another difference between Proxima b and Earth: while Earth's spin period 
is much shorter than its orbital period, Proxima b's rotation has been affected by 
tidal interactions with its host star. The planet is likely to be in one of two 
resonant spin states (see Sect. \ref{sub:spin}).   

In this paper we focus on the evolution of Proxima b's volatile inventory using all 
available information regarding the irradiation of the planet over its lifetime and taking 
into account how tides have affected the planet's orbital and spin evolution. More 
specifically, we address the following issues:
\begin{enumerate}
\item[-] We first estimate the initial water content of the planet by discussing the 
important mechanisms for water delivery occurring in the protoplanetary disk (Sect. 
\ref{init_water}).
\item[-] To estimate the atmospheric loss rates, we need to know the spectrum of Proxima 
at wavelengths that photolyse water (FUV, H Ly$\alpha$) and heat the upper atmosphere, 
powering the escape (soft X-rays and EUV), as well as its stellar wind properties.
For such purpose, we provide measurements of Proxima's high energy emissions and 
wind at the orbital distance of the planet (Sect. \ref{XUV_flux}).
\item[-] To better constrain the system, we investigate the history of Proxima and its planet.
We first reconstruct the evolution of its structural parameters (radius, luminosity), 
the evolution of its high-energy irradiance and particle wind. We then investigate the 
tidal evolution of the system, including the semi-major axis, eccentricity, and rotation period.
This allows us to infer the possible present day rotation states of the planet (Sect. 
\ref{co_evol}).
\item[-] With all the previous information, we can estimate the loss of volatiles of the 
planet, namely the loss of water and the loss of the background atmosphere prior to 
entering in the HZ (the runaway phase) and while in the HZ. To compute the water loss, 
we use an improved energy limited escape formalism \citep{Lammer2003, Selsis2007a} based 
on hydrodynamical simulations \citep{OwenAlvarez2016}. This model was used by 
\citet{Bolmont2016} to estimate the water loss from planets around brown dwarfs and 
the planets of TRAPPIST-1 \citep{Gillon2016} (Sect. \ref{waterloss}).
\end{enumerate}
Following up on the results, in Paper~II we study the possible climate regime that 
can exist on the planet as a function on the volatile reservoirs and the rotation rate 
of the planet.

\begin{table}
\centering
\caption{Adopted stellar and planetary characteristics of the Proxima system.}
\begin{tabular}{lcl}
\hline
Parameter & Value & Source \\
\hline
$M_{\star}$ (M$_{\odot}$)   & 0.123     & This work \\
$R_{\star}$ (R$_{\odot}$)   & 0.141     & \citet{Anglada16}  \\
$L_{\star}$ (L$_{\odot}$)   & 0.00155   & \citet{Anglada16} \\
$T_{\rm eff}$ (K)           & 3050      & \citet{Anglada16} \\
Age (Gyr)                   & 4.8       & \citet{Bazotetal2016} \\
\hline
$M_{\rm p} \sin i$ (M$_{\oplus}$)   & 1.27      & \citet{Anglada16} \\
$a$ (AU)                            & 0.0485    & \citet{Anglada16} \\
$e_{\rm max}$                       & 0.35      & \citet{Anglada16} \\
$S_p$ (S$_{\oplus})$                & 0.65      & \citet{Anglada16} \\
\hline
\end{tabular} 
\label{Char_prox} 
\end{table}

\section{The initial water inventory on Proxima b}\label{init_water}

Proxima b's primordial water content is essential for evaluating the
planet's habitability as well as its water loss and, thus, its present-day
water content. One can easily imagine an unlucky planet located in the
habitable zone that is completely dry, and such situations do arise in
simulations of planet formation \citep{raymond04}. Of course, Earth's
water content is poorly constrained. Earth's surface water budget is 
$\sim$1.5$\times10^{24}$ g, defined as one ``ocean'' of water. The water 
abundance of Earth's interior is not well known. Estimates for the amount 
of water locked in the mantle range between $\lesssim$0.3 and 10 oceans 
\citep{lecuyer98,marty12,panero16}. The core is not thought to contain a
significant amount of hydrogen \citep[e.g.,][]{badro14}.  

In this section we discuss factors that may have played a role in determining
the planet's water content. Our discussion is centered on theoretical 
arguments based on our current understanding of planet formation.  

It is thought that Earth's water was delivered by impacts with water-rich
bodies. In the Solar System, the division between dry inner material and more
distant hydrated bodies is located in the asteroid belt, at $\sim$2.7 AU, 
which roughly divides S-types and C-types \citep{gradie82,demeo13}. 
Earth's D/H and $^{15}$N/$^{14}$N ratios are a match to carbonaceous
chondrite meteorites \citep{marty06} associated with C-type asteroids
in the outer main belt. Primordial C-type bodies are the leading candidate for
Earth's water supply.\footnote{Two Jupiter-family comets and one Oort cloud comet have been measured to
have Earth-like D/H ratios \citep{hartogh11,lis13,biver16} although the Jupiter-family comet 67P/Churyumov–Gerasimenko has a D/H ratio three times higher than Earth's \citep{altwegg15}.
Jupiter-family comets also do not match Earth's $^{15}$N/$^{14}$N ratio 
\citep[e.g.][]{marty16}.}

Models of terrestrial planet formation \citep[see][for recent reviews]{morby12, 
raymond14} propose that Earth's water was delivered by
impacts from primordial C-type bodies. In the classical model of accretion,
water-rich planetesimals originated in the outer asteroid belt 
\citep{morby00,raymond07a,raymond09,izidoro15}. Earth's feeding 
zone was several AU wide and encompassed the entire inner Solar System 
\citep[see fig 3 from][]{raymond06}. In the newer Grand Tack model, water was 
delivered to Earth by C-type material, but those bodies actually condensed 
much farther from the Sun and were both implanted into the asteroid belt 
and scattered to the terrestrial planet-forming region
during Jupiter's orbital migration \citep{walsh11,obrien14}. 

If the Proxima system formed by in-situ growth like our own terrestrial
planets, then there are reasons to think that planet b might be drier than
Earth. First, the snow line is farther away from the habitable zone around
low-mass stars \citep{lissauer07,mulders15}. Viscous heating is the main heat source for 
the inner parts of protoplanetary disks. The location of the snow line is 
therefore determined not by the star but by the disk. However, the location of
the habitable zone is linked to the stellar flux. Thus, while Proxima's
habitable zone is much closer-in than the Sun's, its snow line was likely
located at a similar distance. Water-rich material thus had a far greater
dynamical path to travel to reach Proxima b, and, as expected, water
delivery is less efficient at large dynamical separations \citep{raymond04}. 
But protoplanetary disks are not static. They cool as the bulk of their
mass is accreted by the star. The snow line therefore moves inward in time, 
\citep[e.g.,][]{lecar06,kennedy08,podolak10,martin12}. 
Models for the Sun's early evolution suggest that the Solar 
System's snow line may have spent time as close in as 1 AU 
\citep[e.g.,][]{sasselov00,garaud07}. Yet the Solar System interior to 2.7 AU is 
extremely dry. One explanation for this apparent contradiction is that the 
inward drift of water-rich bodies was blocked when Jupiter formed \citep{morby16}. 
The dry/wet boundary at 2.7 AU may be a fossil remnant of the 
position of the snow line at the time of Jupiter's growth. One can imagine 
that in systems without a Jupiter, such as the Proxima system\footnote{As concluded 
by the UVES M-dwarf survey \citep{Zechmeisteretal2009}, it is highly unlikely that 
Proxima hosts a gas giant within a few AU. Even at longer orbital distances, such 
planet would cause acceleration that would likely be detectable with radial velocity 
data. The Doppler method is only sensitive to the orbital motions along the line-of-sight, 
so there is always a chance that a large planet is hidden to Doppler detection on a 
face-on orbit. As a rule of thumb, we estimate the chance of hiding a gas giant within 
10 AU at $<$10\%. As reported in \cite{Anglada16} there is unconfirmed evidence for 
an additional planet exterior to Proxima b but stellar activity might still be the cause 
of the observed Doppler variability. Even if such a planet is confirmed in the future, 
its minimum mass would be in the range of $\sim$3--6 M$_{\oplus}$. The presence of such 
planet would likely have an impact on the evolution of the putative atmosphere and state 
of Proxima b, especially due to the induction of a non-zero eccentricity and resulting 
non-trivial tidal state \citep{vanLaerhovenetal2014}. However, we consider that adding 
an additional planet and all the associated degrees of freedom to an already complex 
model is an unnecessary complication given the limited information we have on the system 
and the tentative nature of this additional companion. Proxima is possibly the star on 
which the Gaia space mission has highest sensitivity to small planets (Neptune-mass 
objects should be trivially detectable for $P>100$ days irrespective of the inclination), 
so it will not be long before the question about the presence or absence of long-period 
gas giants is finally settled.}, the situation might be quite different. In principle, 
if the snow line swept all the way in to the habitable zone, it may have snowed on 
Proxima b late in the disk's lifetime. 

Second, the impacts involved in building Proxima b were more energetic than
those that built the Earth. The collision speed between two objects in orbit
scales with the local velocity dispersion (as well as the two bodies' mutual
escape speed). The random velocities for a planet located in the habitable 
zone are directly linked to the local orbital speed as $v_{random} \sim
\left(M_\star/r_{HZ}\right)^{1/2}$, where $M_\star$ is the stellar mass
\citep{lissauer07}. For Proxima, the impacts that built planets in the
habitable zone would have been a few times more energetic on average than 
those that built the Earth. This may have led to significant loss of the 
planet's atmosphere and putative oceans \citep{genda05}.  

Third, Proxima b likely took less time than Earth to grow. Assuming a 
surface density large enough to form an Earth-mass planet, simple scaling laws
and N-body simulations show that planets in the habitable zones of 
$\sim$0.1~M$_\odot$ stars form in 0.1 to a few Myr \citep{raymond07b,lissauer07}. 
Even if the planets formed very quickly, the dissipation of 
the gaseous disk after a few Myr \citep{haisch01,pascucci09} 
may have triggered a final but short-lived phase of giant collisions. Although 
it remains to be demonstrated quantitatively, the concentration of impact 
energy in a much shorter time than Earth may have contributed to increased 
water loss.  

Yet simulations have shown that in-situ growth can indeed deliver water-rich
material in to the habitable zones of low-mass stars
\citep{raymond07b,ogihara09,montgomery09,hansen15,ciesla15}. However, these
simulations did not focus on very low-mass stars such as Proxima.  The
water-depleting effects discussed above are expected to increase in importance
for the lowest-mass stars, so the retention of water remains in question.

It also remains a strong possibility that Proxima b formed farther from the
star and migrated inward. Bodies more massive than $\sim$0.1--1~M$_\oplus$ are
subject to migration from tidal interactions with the protoplanetary disk
\citep{goldreich80,paardekooper11}. Given that the mass of
Proxima b is on this order, migration is a plausible origin. Indeed, the
population of ``hot super-Earths'' can be explained if planetary embryos 
formed at several AU, migrated inward to the inner edge of the protoplanetary 
disk and underwent a late phase of collisions 
\citep{terquem07,ida10,mcneil10,swift13,cossou14,izidoro16}. 
If Proxima b or its building blocks formed much farther 
out and migrated inward, then their compositions may not reflect the local 
conditions in the disk. Rather, they could be extremely water-rich 
\citep{kuchner03,leger04}.  If migration did indeed take place it must have
happened very early, during the gaseous disk phase.  Migration would not have
affected the planet's irradiation or tidal evolution, just its initial water
budget.

Other mechanisms may have affected Proxima b's water budget throughout the planet's 
formation.  For example, if Proxima's protoplanetary disk underwent external 
photoevaporation, the snow line may have stayed far from the star 
\citep{kalyaan16}, thus inhibiting water delivery to Proxima b. The short-lived
radionuclide $^{26}$Al is thought to play a vital role in determining the
thermal structure and water contents of planetesimals, especially those that
accrete quickly \citep[as may have been the case for Proxima b's building 
blocks; e.g.,][]{grimm93,desch04}. Finally, we cannot rule
out a late bombardment of water-rich material on Proxima b, although it
would have to have been 1--2 orders of magnitude more abundant than the Solar
System's late heavy bombardment to have delivered an ocean's worth of water
\citep{gomes05}.  

To summarize, there are several mechanisms by which water may have been
delivered to Proxima b. Yet it is unclear how much water would have been
delivered or retained. We can imagine a planet with Earth-like water content
that was delivered somewhat more water than Earth but lost a higher fraction.
We can also picture an ocean-covered planet whose building blocks condensed
beyond the snow line. Finally, we can imagine a dry world whose surface water
was removed by impacts and early heating. In the following sections, we
therefore consider a broad range of initial water contents for Proxima b.

\section{High-energy irradiation} \label{XUV_flux}

High-energy emissions and particle winds have been shown to play a key role in 
shaping the atmospheres of rocky planets. Numerous studies 
\citep[e.g.,][]{Lammeretal2009} have highlighted the impact of the so-called 
XUV flux on the volatile inventory of a planet, including water. The XUV range 
includes emissions from the X-rays (starting at $\sim$0.5~nm -- 2.5 keV) out to 
the far-UV (FUV) just short of the H Ly$\alpha$ line. Here we extend our analysis 
out to 170 nm, which is a relevant interval for photochemical studies.

One unavoidable complication of estimating the XUV fluxes is related to their 
intrinsic variability. Proxima is a well-known flare star 
\citep[e.g.,][]{Haischetal1983,Guedeletal2004,Fuhrmeisteretal2011} and thus 
its high-energy emissions are subject to strong variations (of up to 2 orders
of magnitude in X-rays) over timescales of a few hours and longer. 
Further, optical photometry of Proxima indicates a long-term activity cycle 
of $\sim$7.1 yr \citep{EngleGuinan2011}. For a nearby planet, 
both the so-called quiescent activity and the flare rate of Proxima are 
relevant. X-ray emission of Proxima was observed with ROSAT and XMM. 
\citet{Hunschetal1999} report $\log{L_{\rm X}} = 27.2$~erg~s$^{-1}$ from a 
ROSAT observation, and \citet{SchmittLiefke2004} report 
$\log{L_{\rm X}} = 26.9$~erg~s$^{-1}$ for ROSAT PSPC and 
$\log{L_{\rm X}} = 27.4$~erg~s$^{-1}$ for an XMM observation. It is interesting
to note that Proxima's X-ray flux is quite similar to the solar one, which is 
between $\log{L_{\rm X}} = 26.4$ and 27.7~erg~s$^{-1}$, corresponding to solar 
minimum and maximum, respectively. 

In the present study we estimate the average XUV luminosity over a relatively long 
timescale in an attempt to measure the overall dose on the planetary atmosphere, 
including the flare contribution. This is based on the assumption of a linear 
response of the atmosphere to different amounts of XUV radiation, which is certainly 
an oversimplification, but should be adequate for an approximate evaluation
of volatile loss processes.

High-energy observations of Proxima have been obtained from various facilities
and covering different wavelength intervals. In the X-ray range we use
XMM-Newton observations with Observation IDs 0049350101, 0551120201,
0551120301, and 0551120401. The first dataset, with a duration of 67 ks, was
studied by \citet{Guedeletal2004} and contains a very strong flare with a total
energy of $\approx2\times10^{32}$ erg. The other three (adding to a total of
88 ks), were studied by \citet{Fuhrmeisteretal2011}, and include several
flares, the strongest of which has an energy of about $2\times10^{31}$ erg.  

The flare distribution of Proxima can be crudely approximated using the analysis 
of \citet{Audardetal2000} for CN Leo, which has similar X-ray luminosity and 
spectral type. \citet{Audardetal2000} find a cumulative flare
distribution of CN Leo that can be described by a power law with the form
$N(>E)=3.7\times10^{37} E^{-1.2}$, where $N$ is the number of flares per day,
and $E$ is the total (integrated) flare energy in erg. Thus, CN~Leo has flares
with energies greater than about $2\times10^{31}$ erg over a timescale of 1 
day. Interestingly, this is in agreement with the 88-ks dataset of Proxima, and 
thus this seems to be quite representative of the daily average X-ray flux. 
The (time-integrated) average flux from the XMM 88-ks dataset between 
0.65 and 3.8 nm yields a value at the orbital distance of Proxima b of 
87 erg s$^{-1}$ cm$^{-2}$.

\begin{figure}
\centering
\includegraphics[width=\linewidth]{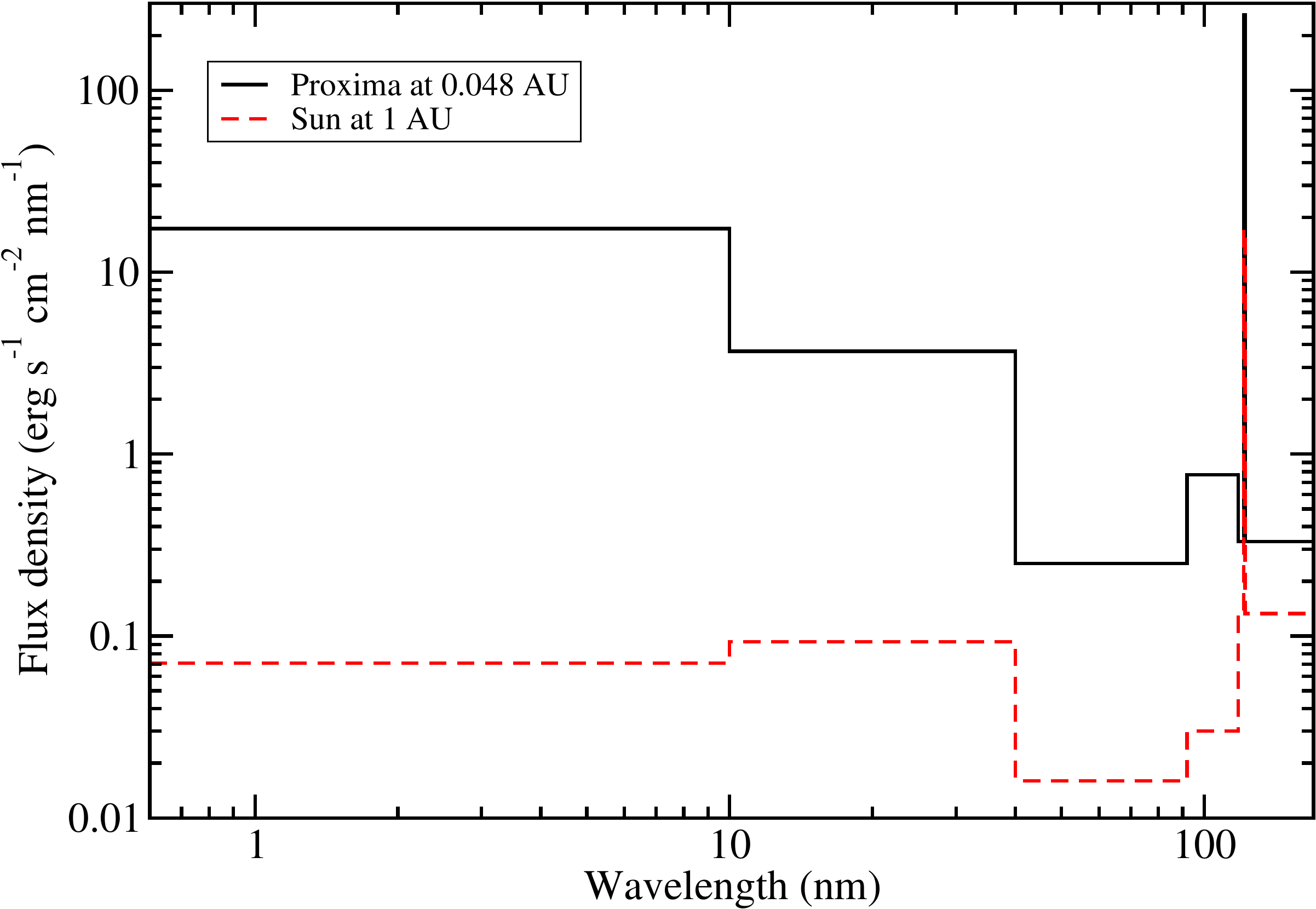}
\caption{High-energy spectral irradiance received by Proxima b and the Earth. The values 
correspond to those in Table \ref{tab:XUV} but calculated per unit wavelength (i.e., 
divided by the width of the wavelength bin; 0.5 nm is adopted for H Ly$\alpha$).}
\label{fig:XUVspec}
\end{figure}

Using the expressions in \citet{Audardetal2000} we can estimate a correction factor
to account for the total energy produced by more energetic flares. The integrated
flux value of 87 erg s$^{-1}$ cm$^{-2}$ should represent the average flux between 
energies of $2\times10^{29}$~erg \citep[minimum energy as found by][]{Audardetal2000} 
and $2\times10^{31}$~erg, and should be compared with the flux produced by flares up 
to $2\times10^{32}$~erg, which is the strongest flare observed for Proxima. The 
integration of the cumulative flare distribution above indicates that the X-ray dose 
produced by energetic flares increases the typical 1-day average by about 25\%. 
Thus, this implies an extra flux of 22~erg~s$^{-1}$~cm$^{-2}$, and a total average 
of 109~erg~s$^{-1}$~cm$^{-2}$ with the energetic flare correction. This value and those 
following in this section are listed in Table \ref{tab:XUV} and illustrated in 
Fig. \ref{fig:XUVspec}.

Note that a comparison of the results in \citet{Walker1981} and 
\citet{Kunkel1973} indicates the Proxima has about 60\% of the flare rate of 
CN Leo as measured in comparable energy bands. However, this small difference 
does not affect our calculations since we are interested in estimating the 
relative contribution of the energetic flares with respect to the background of
lower energy flare events. The methodology assumes a power law slope as given 
above for CN Leo (and should apply to Proxima as well) and some flare energy 
intervals that are appropriate for Proxima. 

ROSAT observations were used in the wavelength range from 3.8 to 10 nm. 
Four suitable datasets are available from the ROSAT archive, with Dataset IDs
RP200502A01, RP200502A02, RP200502A03, and RP200502N00, and integration times
ranging from 3.8 to 20 ks. After flare events were filtered out, quiescent
fluxes were calculated by fitting a two temperature (2-T) MEKAL collisional
ionization equilibrium model \citep{Drakeetal1996} with solar abundance 
\citep{Nevesetal2013} and $N_H$ value of $4\times10^{17}$~cm$^{-2}$. This was 
done within the XSPEC (v11) X-ray Spectral Fitting Package, distributed by 
NASA’s HEASARC. Because of the short integration times, substantial 
differences between datasets exist depending on the flare properties.  We 
employed the RP200502N00 dataset because it has a 0.6--3.8~nm integrated flux 
closer to the XMM values (and this ensured similar spectral hardness). A 
modest scaling of 1.28 was used to bring the actual fluxes into agreement, 
including the flare correction. Using this prescription, we calculated that 
the 3.8--10~nm flux of Proxima at the distance of its planet is 
43~erg~s$^{-1}$~cm$^{-2}$ in quiescence and 54~erg~s$^{-1}$~cm$^{-2}$ with 
the flare correction. Thus, the total X-ray dose of Proxima b from 0.6 to 10~nm 
is 163~erg~s$^{-1}$~cm$^{-2}$, including an energetic flare correction of 
33~erg~s$^{-1}$~cm$^{-2}$.

An alternative approach to estimate the flare-corrected X-ray flux of Proxima
is to use the similarity with the Sun. Proxima's cumulative energy
distribution can be compared to the solar and other stellar distributions in
\citet{Drakeetal2015}. The cumulative flare energy output of the quiet Sun is
$2\times10^{25}$~erg~s$^{-1}$ \citep{Hudson1991}, i.e., 3.1~erg~s$^{-1}$~cm$^{-2}$ 
at the distance of Proxima b. The current Sun, as well as average Sun-like stars 
observed by Kepler, and Proxima, are flaring at roughly the same rate, which is a 
factor of $\sim$10 higher than solar minimum \citep{Shibayamaetal2013}. 
This results in a flux of about 31~erg~s$^{-1}$~cm$^{-2}$ at the distance of 
Proxima b, nicely consistent with the estimate above.

For the extreme-UV range we use the EUVE spectrum available from the mission 
archive with Data ID proxima\_cen\_\_9305211911N, corresponding to an 
integration time of 77 ks. This dataset was studied by \citet{Linskyetal2014},
who measured an integrated (and corrected for interstellar medium -- ISM -- 
absorption) flux between 10 and 40~nm of 89~erg~s$^{-1}$~cm$^{-2}$ at the 
distance of Proxima b. No information on the flare status of the target is 
available, and we applied the same correction obtained for the X-rays to 
obtain a flux value of 111~erg~s$^{-1}$~cm$^{-2}$ at the distance 
of Proxima b. 

FUSE observations are used to obtain the flux in part of the far-UV range. 
We employed the spectrum with Data ID D1220101000 with a total integration
time of 45 ks. A concern related to FUSE observations is the contamination
by geocoronal emission. We made sure that the spectrum had little or no 
visible geocoronal features but the H Ly lines always show some degree of
contamination. Such flux increase competes with the significant ISM 
absorption, which diminishes the intrinsic stellar flux. We measured the 
integrated flux in the 92--118~nm interval excluding the H Ly series and 
obtained 10~erg~s$^{-1}$~cm$^{-2}$. Following \citet{Guinanetal2003} and 
\citet{Linskyetal2014}, we estimate the H Ly series contribution, except H 
Ly$\alpha$, to be of the same order and thus the 92--118~nm flux at the 
distance of Proxima b is $\approx$20~erg~s$^{-1}$~cm$^{-2}$. Note that 
\citet{Christianetal2004} found 3 flare events in the FUSE dataset, which 
produce an increase of up to one order of magnitude in the instantaneous flux. 
The integrated effect of such flares is about 20--30\% relative to the 
quiescent emission, which appears to be reasonable given our X-ray estimates 
and thus no further correction was applied.

A high-quality HST/STIS spectrum obtained from the StarCAT catalog
\citep{Ayres2010} was used to estimate the fluxes between 118 and 170~nm 
(except for H Ly$\alpha$). The flux integration yielded a value of 
17~erg~s$^{-1}$~cm$^{-2}$ at the distance of Proxima b. A flare analysis 
of this dataset was carried out by \citet{LoydFrance2014}, who identified 
a number of flare events in the stronger emission lines. These flares 
contribute some 25--40\% of the integrated flux (Loyd, priv. comm.) and 
thus represent similar values to those found in the X-ray domain. No further 
corrections were made. The same base spectrum was used by \citet{Woodetal2005} 
to estimate the intrinsic H Ly$\alpha$ stellar line profile and the integration 
results in a flux of 130~erg~s$^{-1}$~cm$^{-2}$ at the orbital distance of 
Proxima b. The relative flare contribution corrected for ISM absorption is 
estimated to be of $\sim$10\% (Loyd, priv. comm.). 

The interval between 40 and 92~nm cannot be observed from Earth due to the 
very strong ISM absorption, even for a star as nearby as Proxima. To estimate
the flux in this wavelength range we make use of the theoretical calculations
presented by \citet{Linskyetal2014}, who show that it can be approximated as
being about 10\% of the H Ly$\alpha$ flux, i.e., 13~erg~s$^{-1}$~cm$^{-2}$.

Thus, the total integrated flux today that is representative of the
time-averaged high-energy radiation on the atmosphere of Proxima b is of 307
erg~s$^{-1}$~cm$^{-2}$ between 0.6 and 118~nm. To compare with the current 
Earth XUV irradiation we employ the \citet{Thuillieretal2004} solar spectrum
corresponding to medium solar activity and the average of the maximum and 
minimum Solar Irradiance Reference Spectrum (SIRS) as given by 
\citet{Linskyetal2014}. Both data sources provide very similar results. 
Integration in the relevant wavelength interval yields a total XUV flux at 
Earth of 5.1~erg~s$^{-1}$~cm$^{-2}$. Thus, Proxima b receives 60 times more XUV
flux than the current Earth, which we refer to as XUV$_\oplus$. Also, the far-UV 
flux on Proxima~b between 118~nm and 170~nm is 147~erg~s$^{-1}$~cm$^{-2}$, which is 
about 10 times higher than the flux received by the Earth, namely FUV$_\oplus$. 
The H Ly$\alpha$ flux alone received by Proxima is 15 times stronger than Earth's. 
Note that the high-energy emission spectrum of Proxima is significantly harder 
than that of the Sun today. If we consider that the current X-ray luminosities of the 
Sun and Proxima are similar, the distance scaling from 1 AU to 0.048 AU represents 
a factor of 435 in the flux, which is much higher than our measured value of 60. 
All the values measured for Proxima as well as the comparison with the Sun are 
listed in Table \ref{tab:XUV} and illustrated in Fig. \ref{fig:XUVspec}.

\begin{table}[t]
\centering
\begin{tabular}{lccr}
\hline
Wavelength interval (nm) & Proxima b & Earth & Ratio \\
\hline
0.6--10 (X-rays)      &  163  &  0.67   & $\approx$250 \\
10--40                &  111  &  2.8    & $\approx$40 \\
40--92                &   13  &  0.84   & $\approx$15 \\
92--118               &   20  &  0.79   & $\approx$25 \\
0.6--118 (XUV)        &  307  &  5.1    & $\approx$60 \\
10--118 (EUV)         &  144  &  4.4    & $\approx$30 \\
118--170 (FUV)        &  147  &  15.5   & $\approx$10 \\
H Ly$\alpha$ (122 nm) &  130  &  8.6    & $\approx$15 \\
\hline
\end{tabular}
\caption{High-energy fluxes received currently by Proxima b and the Earth
in units of erg~s$^{-1}$~cm$^{-2}$.}
\label{tab:XUV}
\end{table}

\section{Co-evolution of Proxima~b and its host star}
\label{co_evol}

The observations of the Proxima system \citep{Anglada16} show that Proxima b is
located in the classical insolation HZ \citep[as defined 
in][]{Kasting1993, Selsis2007, Kopparapu2013, Kopparapu2014}. However, as Proxima 
is a low-mass star, it spent a non-negligible time decreasing its luminosity during 
the early evolution, which means that the HZ moved inwards with time. 
If Proxima~b's orbit remained the same with time, and assuming it was formed with a
non-zero water reservoir, it would have experienced a runaway greenhouse 
phase, which means water was in gaseous phase prior to entering the HZ. 
Planets orbiting very low-mass stars could be desiccated by this hot 
early phase and enter the HZ as dry worlds 
\citep[as shown by the works of][]{BarnesHeller2013, LugerBarnes2015}. 
In contrast, the detailed analysis of the TRAPPIST-1 system \citep{Gillon2016} by 
\citet{Bolmont2016}, using a mixture of energy-limited escape formalism together 
with hydrodynamical simulations \citep{OwenAlvarez2016}, shows that the planets 
could have retained their water during the runaway phase. We apply a similar scheme 
to Proxima b to evaluate this early water loss.

\subsection{The early evolution of Proxima}\label{evolutionary_model}

\begin{figure}
\begin{center}
\includegraphics[width=9cm]{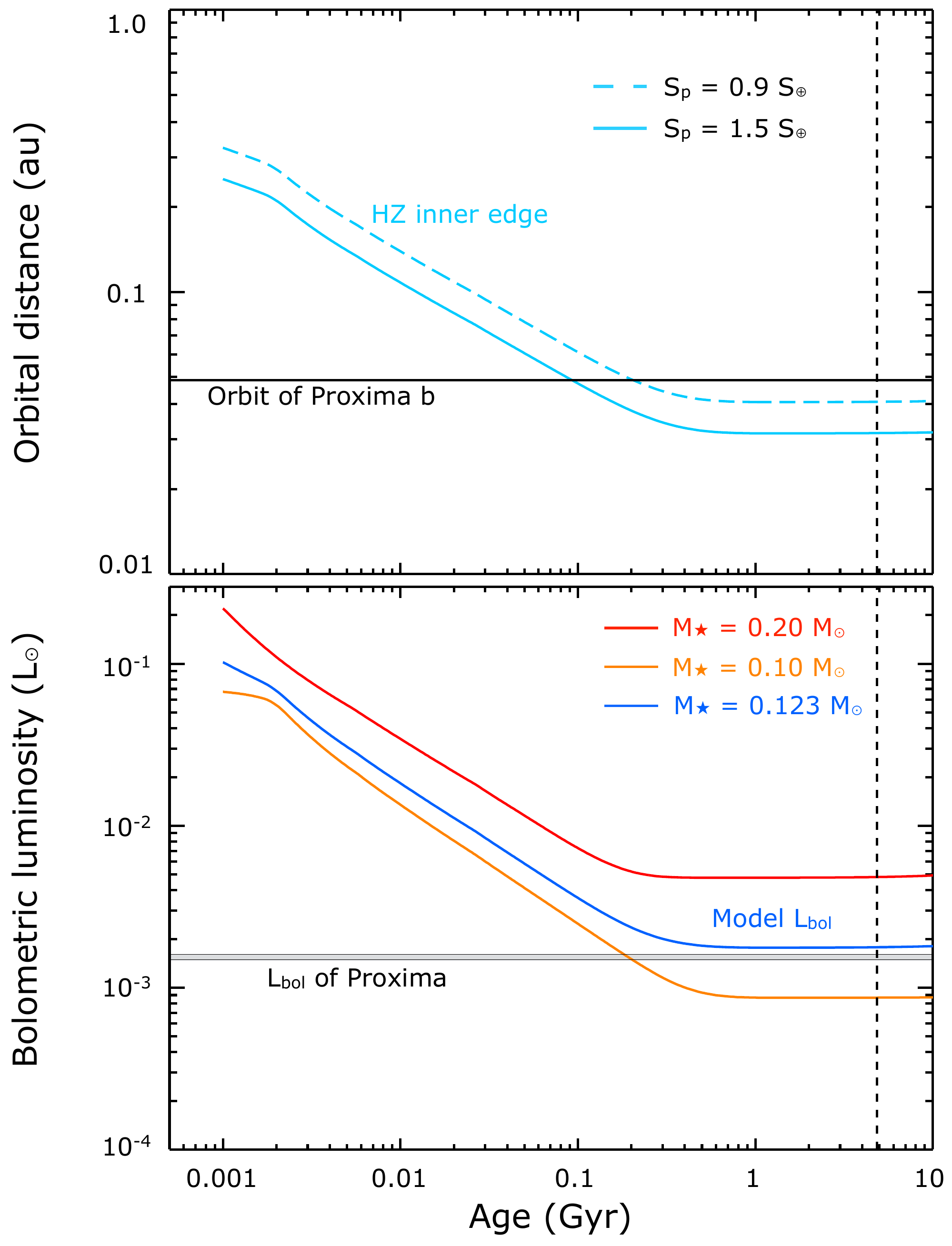}
\caption{Evolution of the HZ inner edge, bolometric luminosity and XUV luminosity for 
Proxima. Top panel: evolution of the inner edge of the HZ for two different 
assumptions: $S_p = 0.9$~S$_\oplus$ (dashed blue line), $S_p = 1.5$~S$_\oplus$ (full 
blue line). The full black line corresponds to Proxima's measured orbital distance. 
Bottom panel: evolution of the luminosity for a $0.1$~M$_\odot$ star (in orange), for 
0.2~M$_\odot$ (in red) and 0.123~M$_\odot$ (in blue). The grey area corresponds to 
the observed value (see Table \ref{Char_prox}). The black vertical dashed line 
corresponds to the estimated age of Proxima.}
\label{Prox_HZ_lumi_0123_Msun}
\end{center}
\end{figure}

Proxima's physical properties, such as its mass, radius, luminosity and effective
temperature are given in Table \ref{Char_prox}. We used the evolutionary tracks
provided by \citet{Baraffe2015} in order to reproduce these values at the age
of the star (4.8~Gyr). As $M_{\star}=0.123$~M$_{\odot}$ is not tabulated, we 
performed a linear interpolation between the evolutionary tracks corresponding to 
0.1~M$_{\odot}$ and 0.2~M$_{\odot}$. We tested the following masses: 0.120, 
0.123, 0.125, 0.130~M$_{\odot}$. None of these interpolated tracks allow to 
reproduce simultaneously the exact values of the adopted radius, luminosity and 
effective temperature simultaneously. Thee best agreement for the luminosity is found 
for a mass of 0.120~M$_{\odot}$ but the best agreement for the effective temperature 
is found for a mass of 0.130~M$_{\odot}$. For the radius, all masses lead to an 
agreement. This apparent (minor) disagreement between luminosity, effective temperature 
and mass may come from the fact that the models of \citet{Baraffe2015} use a solar 
metallicity while Proxima is more metal rich than the Sun ($[Fe/H]$ = 0.21, 
\citealt{Anglada16}). In the following we assume a mass of 0.123~M$_{\odot}$. 
Fig. \ref{Prox_HZ_lumi_0123_Msun} shows the evolution of the bolometric
luminosity of Proxima, according to our adopted model.

To estimate the location of the inner edge of the HZ, we considered two possible scenarios 
for the rotation of the planet, as discussed in section~\ref{sub:spin}: a synchronous 
rotation and a 3:2 spin-orbit resonance. For a non-synchronous planet we considered an 
inner edge at $S_{\rm p}=0.9$~S$_{\oplus}$, where S$_{\oplus}=1366$~W~m$^{-2}$ is the 
flux received by the Earth (e.g., \citealt{Kopparapu2013, Kopparapu2014}). For a 
synchronized planet, we locate the inner edge at $S_{\rm p}=1.5$~S$_{\oplus}$ 
(the protection of the substellar point by clouds allows the planet to be much 
closer, e.g., \citealt{Yang2013, Kopparapu2016}). The top panel of Fig. 
\ref{Prox_HZ_lumi_0123_Msun} shows the evolution of the inner edge of the HZ for 
both prescriptions compared to the semi-major axis of Proxima b.

\subsection{History of XUV irradiance}\label{irradiation}

In addition to the average flux that Proxima b receives today given in Section \ref{XUV_flux}, 
having an approximate description of the history of XUV emissions is key to investigate the 
current atmospheric properties of the planet and its potential habitability. While the variation 
of XUV emissions with time is relatively well constrained for Sun-like stars 
\citep{Ribasetal2005,Claireetal2012}, the situation for M dwarfs (and especially mid-late M 
dwarfs) is far from understood. Some results were presented and discussed by \citet{Selsis2007}
and, more recently, by \citet{Guinanetal2016} within the ``Living with a Red Dwarf'' program. 
Qualitatively, these works present a picture of a time-evolution in which 
$\log L_{\rm X}/L_{\rm bol}$ shows a flat regime starting at the ZAMS and extending out to 
about 1 or a few Gyr, and known as saturation \citep[e.g.,][]{JardineUnruh1999}, followed 
by a regime in which the decrease shows a power law form. The timescales in this 
approximation are notoriously uncertain. For example, models of
low-mass star angular momentum evolution predict that braking timescales in low-mass stars 
are substantially longer than in Sun-like stars. \citet{Reiners12} estimate a timescale of 
roughly 7 Gyr until activity in a star like Proxima falls below the saturation limit, which is 
often assumed to be around a few tens of days. From there, the star would need a few more Gyr 
to reach the observed value of $P = 83$ d. This would imply an age of about 10 Gyr or so, which 
is clearly inconsistent with the estimate of $4.8\pm1$ Gyr by \citet{Bazotetal2016}. A way out 
is that Proxima started its rotational evolution with less initial angular momentum or was kept 
at fixed rotation rate by a surrounding disk for longer than the canonical 10 Myr. The 
saturation limit itself is not well constrained in stars of such low masses. For example, 
extrapolating the luminosity scaling law from \citet[][Eq.\,10]{Reinersetal2014}, the 
saturation limit would be at $P_{\rm sat} \approx 40$\,d. However, the radius-luminosity 
relation used in that paper cannot readily be extended to very low masses and a calculation 
using their saturation criterion yields $P_{\rm sat} \approx 80$ d. The latter would imply 
that Proxima is still exhibiting saturated activity and probably did so over its entire 
lifetime. \cite{Reinersetal2014} also suggest that the amount of X-ray emission may slightly 
depend on $P$ in saturated stars such that Proxima would have had a higher value of $L_{\rm 
X}/L_{\rm bol}$ when it was young.  

\begin{figure}
\centering
\includegraphics[width=\linewidth]{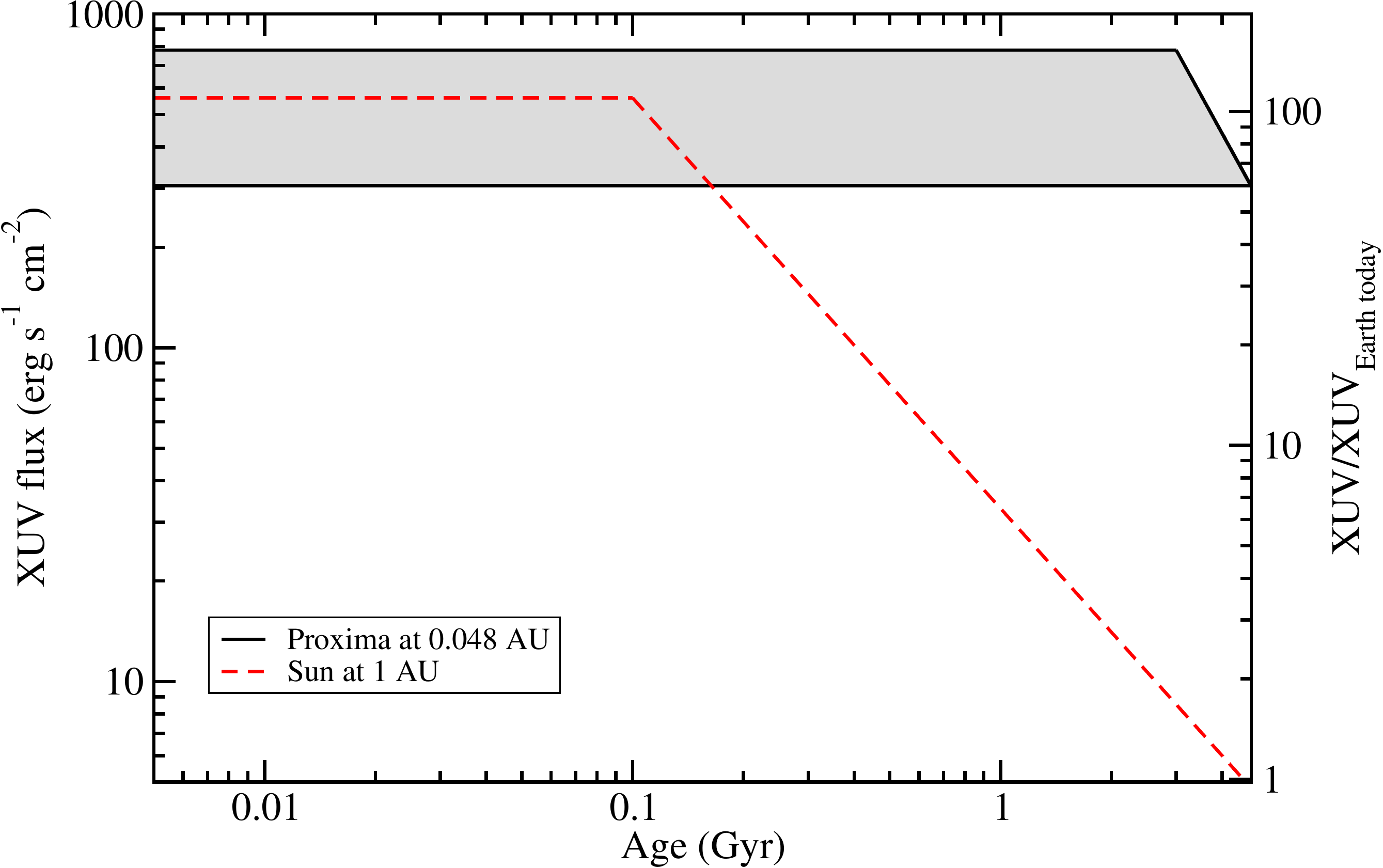}
\caption{XUV flux evolution for Proxima and the Sun at the orbital distance
of Proxima b and Earth, respectively. The two scenarios discussed in the text, 
namely a flat regime and a power-law decrease and a constant value throughout,
are represented, with the gray area indicating a realistic possible range. Such 
relationships show that Proxima b has been irradiated at a level significantly 
higher than the Earth throughout most of their lifetimes and the integrated XUV 
dose is between 7 and 16 times higher, depending on the assumed XUV flux evolution 
for Proxima.}
\label{fig:XUVevol}
\end{figure}

Clearly, there are significant uncertainties in low-mass star angular momentum evolution. In 
any case, all models and observations of rotational braking tend to agree that stars like 
Proxima exhibit saturated activity from early ages until an age of several Gyr, perhaps even 
until today. If today's rotation period is below the saturation limit, exponential rotational 
braking on timescales of Gyr and $L_{\rm X} \propto P^{-2}$ is expected. We calculate two 
scenarios to estimate the effect of angular momentum evolution on the history of XUV irradiance. 
In the first scenario, we estimate that Proxima stayed saturated at a level of $\log L_{\rm
X}/L_{\rm bol} =-3.3$ until an age of 3 Gyr and spent another 2 Gyr until its rotation 
decreased to 83 d as observed (both values with uncertainties of about 1 Gyr) and its X-ray 
radiation diminished to $\log L_{\rm X}/L_{\rm bol} =-3.8$ as observed in quiescence. Since 
in this scenario the spectral hardness of the high-energy emissions has likely decreased 
with time \citep[as happens for Sun-like stars; see][]{Ribasetal2005}, we adopt an 
approximate slope of $-2$ for the XUV and we suggest the following relationship:
\begin{eqnarray}
&&F_{\rm XUV}=7.8\times10^2 \hspace*{20mm} \mbox{for $\tau < \tau_\circ$}\nonumber\\
&&F_{\rm XUV}=7.8\times10^2 \: [\tau/\tau_\circ]^{-2} \hspace*{7.5mm} \mbox{for $\tau > \tau_\circ$}
\end{eqnarray}
with $\tau_\circ = 3$ Gyr and $F_{\rm XUV}$ (0.6--118 nm) in erg~s$^{-1}$~cm$^{-2}$ 
at 0.048 AU. Thus, in our first scenario, Proxima b was probably irradiated by XUV photons at 
a level $\sim$150 XUV$_\oplus$ during the first 3 Gyr of its lifetime. This functional 
relationship is shown in Fig. \ref{fig:XUVevol}. We provide a second scenario in which we 
assume that Proxima has remained in a saturated activity state for its entire lifetime and 
that the saturation level is the same as the one observed today, $\log L_{\rm X}/L_{\rm bol} 
=-3.8$. This is clearly a lower limit to the XUV radiation emitted by Proxima and it is 
also shown in Fig. \ref{fig:XUVevol}.

We can also compare the integrated XUV irradiance that Proxima b and the Earth have 
likely received over the course of their lifetimes. For the Earth and the Sun, we use 
the expressions in \citet{Ribasetal2005} with a slight correction to reflect the 
updated XUV current solar irradiance value discussed in Sect. \ref{XUV_flux}. 
This corresponds to $F_{\rm XUV}=5.6\times10^2$~erg~s$^{-1}$~cm$^{-2}$ up to 0.1~Gyr and 
$F_{\rm XUV}=33\:\tau^{-1.23}$~erg~s$^{-1}$~cm$^{-2}$ beyond. The calculations show 
that Proxima b has received, in total, between 7 and 16 times more XUV radiation than 
Earth, with this range corresponding to the two XUV evolution scenarios described above.

\subsection{Particle wind}\label{sub:wind}

The analysis of \citet{Woodetal2001} using the astrospheric absorption of the 
H Ly$\alpha$ feature provided an upper limit of the mass loss rate of Proxima of
0.2~$\dot{\rm M}_\odot$ ($4\times10^{-15}$~M$_\odot$~yr$^{-1}$). For the Sun,
the latitudinal particle flux depends on the activity level, ranging from 
nearly spherically symmetric during solar maximum to being significantly higher
at the ecliptic (Solar) equator with respect to the poles during solar minimum
\citep{Sokoletal2013}. Nothing is known on the geometry of particle emissions for 
other stars. Thus, to scale the mass loss rate from the solar value at Earth to 
the value that Proxima b receives, we adopt two different geometrical prescriptions, 
namely, a spherical distribution (i.e., flux scaling with distance squared) and an 
equatorial distribution (i.e., flux scaling with distance). Following these recipes, 
we estimate that Proxima b is receiving a particle flux that could be within a factor 
of $\approx$4 and $\approx$80 of today's Earth value. Note that the methodology of 
determining mass loss rates from the observation of the astrospheric absorption assumes
a constant or quasi-steady mass loss rate \citep{LinskyWood2014} and should represent 
a time-averaged value, comprising both the quiescent stellar particle emissions and 
coronal mass ejections possibly related to flare events. 

Regarding the history of the particle fluxes, not much is known on the evolution
of stellar wind over time. The observations of \citet{Woodetal2014}, in good 
agreement with the recent semi-empirical analyses of \citet{doNascimentoetal2016} 
and \citet{AirapetianUsmanov2016}, reveal a picture in which the mass loss rate 
of Sun-like stars was quite similar to today's during the early evolution (up to 
about 0.7 Gyr in the case of the Sun), then there is evidence of much stronger wind
fluxes of about 50--100 times today's value, and the subsequent evolution follows
a power-law relationship with age with an exponent of about $-2$. In this picture, 
which is based on the surface X-ray flux, Proxima would still be in the initial 
low-flux regime and therefore the upper limit of 4--80 times today's Earth value can 
be assumed to apply to its entire lifetime. 

\subsection{The magnetopause radius of Proxima b}\label{magnetopause}

In order to provide a first estimate of the position of the magnetopause of Proxima~b we 
follow the approach of \cite{Vidotto13}. This method assumes an equilibrium between 
the magnetic pressures associated with the stellar and planetary magnetic fields. By 
doing so we neglect the effect of the ram pressure of the stellar wind and only consider 
the stellar magnetic pressure -- which \cite{Vidotto13} found to be important for 
M dwarfs -- and we therefore derive an upper limit for the magnetopause radius. 
Starting from the basic equation 2 of \cite{Vidotto13}, which defines the pressure 
balance at the nose of the magnetopause, and rescaling with respect to the Earth case 
(for a balance driven by the wind ram pressure of the stellar wind in that case) we obtain 
the following expression for the radius of the magnetopause relative to the planetary 
radius:
\begin{equation}
  \dfrac{r_M}{r_p} = K 
  \left( \dfrac{R_{orb}}{{\rm [1~au]}} \right)^{2/3} 
  \left( \dfrac{R_\star}{R_\odot} \right)^{-2/3} 
  \left( \dfrac{B_{p,0}}{f_1 f_2 B_\star} \right)^{1/3}   
  \label{eq:rmag_numeric}
\end{equation}
where $K=15.48$, $B_{p,0}$ is the polar magnetic field of the planet taken at the 
surface, $B_\star$ is the average stellar magnetic field taken at photosphere, and 
$f_1$ and $f_2$ are two scaling factors. We use the values $f_1=0.2$ if the 
large-scale component of the stellar magnetic field is dipole-dominated and 
$f_1=0.06$ if it is multipolar. From the stellar sample of \cite{Vidotto13} we 
find $f_2=1/15$ for most stars and larger values for stars with a very non-axisymmetric 
field, we use $f_2=1/50$ as a representative value of this case. The derivation of 
Eq.~(\ref{eq:rmag_numeric}) and the meaning of the factors $f_1$ and $f_2$ are 
detailed in Appendix \ref{app1}. As in \cite{Vidotto13}, we compute $B_\star$ 
using the parametrization of \cite{Reiners12} :
\begin{equation}
\begin{array}{rll}
  B_\star =& B_{\rm crit} & \text{for } P \leq  P_{\rm crit}\\
  B_\star =& B_{\rm crit} \left(\dfrac{P_{\rm crit}}{P}\right)^a & 
    \text{for } P >  P_{\rm crit}\text{  ,}\\
\end{array}
\label{eq:rmag_Bstar}
\end{equation}
where we use $B_{\rm crit}=3~{\rm kG}$ and $a=1.7$ as in \cite{Vidotto13}. In line
with the discussion in Sect.~\ref{irradiation}, we adopt two scenarios for the 
magnetic properties of Proxima. One in which the star is still very near the 
saturation rotation period $P_{\rm crit}$ and another one by which it stayed 
saturated until about 2 Gyr ago, which would roughly correspond to 
$P_{\rm crit} = 40$~d and thus $B_\star \approx 1$~kG.

In order to provide a more realistic estimate, we also take into account the impact of 
the stellar wind ram pressure on the magnetopause radius of Proxima~b. The detail of 
the calculations is provided in Appendix \ref{app1}. From our estimate in 
Sect.~\ref{sub:wind} that the wind particle flux at Proxima is 4--80 times that at 
Earth, it ensues that the ram pressure exerted by the stellar wind of Proxima at 
Proxima~b is also 4--80 times the ram pressure of the solar wind at Earth, if we 
assume that both stars have the same wind velocity (this assumption is used for 
instance in \citealt{Woodetal2001}).  

Assuming a planetary magnetic field equal to the value for the Earth  
$B_{p,0} = B_{p,0}^\oplus$, we derive $r_M/r_p$ values ranging from 2.2 to 6.9.
For a weaker field $B_{p,0} = 0.2~\,B_{p,0}^\oplus$, more in line with \cite{Zuluaga13} for a 
tidally-locked planet, we obtain values ranging from 1.3 to 4.1. These results are 
summarized in Table~\ref{tab:rmag}. In the cases with a dipole-dominated stellar 
magnetic field the wind ram pressure has virtually no impact on the magnetopause 
radius, while for the multipolar stellar magnetic field cases considered we provide 
a range of values for $r_M/r_p$ representing various values for the wind ram pressure 
(varying by a factor of 20). We note that for a star star such as Proxima in the 
low-wind flux regime, even in those cases where the ram pressure is non-negligible, 
the magnetic pressure of the stellar wind remains the dominant term, in contrast 
with the case of the Earth.

\begin{table}
\centering
\caption{Estimates of the size of the magnetopause of Proxima~b derived from Eqs. 
(\ref{eq:rmag_numeric}) and (\ref{eq:rmag_Bstar}). We consider two values for the 
intrinsic magnetic field strength of Proxima~b and  both dipolar and multipolar 
configurations once the star has left the saturated regime. We also consider two 
possible values for the magnetic field of Proxima~b, and a range of values for the 
ram pressure of the stellar wind (see text).}
\begin{tabular}{cccc}
\hline 
$B_\star$  & field    & \multicolumn{2}{c}{$r_M / r_p$} \\
 (G)        & geometry & $B_{p,0}=B_{p,0}^\oplus$ & $B_{p,0}=0.2\, B_{p,0}^\oplus$\\
\hline
$3\times10^3$ & dipolar    &  2.2      & 1.3     \\
$1\times10^3$ & dipolar    &  3.2      & 1.9\\
$1\times10^3$ & multipolar &  5.4--6.9 & 3.2--4.1\\
\hline
\end{tabular}
\label{tab:rmag}
\end{table}

\subsection{Orbital tidal evolution}\label{tid_evol}

As Proxima b is located close to its host star, its orbit is likely to have
suffered tidal evolution. To investigate this possibility, we adopted a standard 
equilibrium tide model \citep{Hut1981, Mignard1979, EKH1998}, taking into account 
the evolution of the host star \citep[as in][]{Bolmont2011, Bolmont2012}. We 
tested different dissipation values for the star: from the dissipation in a Sun-like 
star to the dissipation in a gas giant following \citet{Hansen2010}, which differ by 
several orders of magnitude. 
We assume an Earth composition for the planet, which gives us a radius of 
$\sim$1.1~R$_{\oplus}$ for a mass of $1.3$~M$_{\oplus}$ \citep{Fortney2007} and a resulting 
gravity $g=10.5$~m~s$^{-2}$. Finally, we explored tidal dissipation factors for the planet  
ranging from ten times lower than that of Earth \citep[][hereafter noted 
$\sigma_{p}$]{deSurgyLaskar1997} to the Earth's value. The Earth is thought to be very 
dissipative due to the shallow water reservoirs \citep[as in the bay of 
Biscay,][]{GerkemaLamMaas2004}. In the absence of surface liquid layers, i.e. before 
reaching the HZ, the dissipation of the planet would therefore be smaller than that of the
Earth. Considering this range in tidal dissipation should encompass what we expect for this 
planet.

As in \citet{Bolmont2011} and \citet{Bolmont2012}, we compute the effect of both 
the tide raised by the star on the planet (planetary tide) and by the planet on the 
star (stellar tide). In agreement with \citet{Bolmont2012}, we find that, even 
when assuming a high dissipation in the star, no orbital evolution is induced by the 
stellar tide. The semi-major axis and inclination of the planet remain constant
throughout the evolution and are thus independent of the wind prescription governing 
the spin evolution of the star. The planet is simply too far away.

The planetary tide leads mainly to an evolution of the planet's rotation 
period and obliquity (see Sect. \ref{sub:spin}). 
The eccentricity evolves on much longer timescales so that it does not decrease
significantly over the 4.8~Gyr of evolution. Assuming a dissipation of 0.1$\sigma_p$, 
we can reproduce the observed upper limit eccentricity \citep[0.35, given by][]{Anglada16} 
and semi-major axis at present day with a planet initial semi-major axis and eccentricity 
of $\sim$0.05~AU and $\sim$0.37, respectively. The current eccentricity of 0.35 would imply 
a tidal heat flux in the planet on the order of 2.5~W~m$^{-2}$, i.e. comparable to the 
one of Io \citep{Spencer2000}. This would imply an intense volcanic activity on the 
planet. The average flux received by the planet would also be increased by 6--7\% 
compared to a circular orbit with the same semi-major axis. 

An orbital eccentricity of 0.37 at the end of accretion may be too high. Let us assume 
Proxima~b is alone in the system, its eccentricity could thus be excited only by 
$\alpha$ Centauri. Given the structure of the system \citep{Kaib2013, 
WorthSigurdsson2016}, this excitation should not be responsible for eccentricities 
higher than 0.1. We therefore computed the tidal evolution of the system with an 
initial eccentricity of 0.1. By the age of the system and assuming a dissipation of 
0.1$\sigma_p$, the eccentricity would have decreased to 0.097, and the tidal heat 
flux would be $\sim$0.07~W~m$^{-2}$, which is of the order of the heat flux of the 
Earth \citep{Pollack1993}. Assuming a dissipation as the one of the Earth, we find 
that the eccentricity would have decreased to $0.07$, which corresponds to a tidal 
heat flux of $\sim$0.03~W~m$^{-2}$.

\subsection{Is Proxima b synchronously rotating?}\label{sub:spin}

Although it has been shown that the final dynamical state of an isolated star-planet 
system subjected only to gravitational tides should be a circular orbit and the 
synchronization of both spins \citep{Hut80}, there are several reasons for not finding 
a real system in this end state:
\begin{itemize}
\item[$\bullet$] Tidal evolution timescales may be too long for the system to reach 
equilibrium. In the case at hand, it has indeed been shown above that circularization is 
expected to take longer than the system's lifetime. The spin evolution,
however, is expected to be much faster so that synchronous rotation would be
expected.\footnote{For a slightly eccentric planet, some simple tidal models
predict a slow ``pseudosynchronous rotation,'' whose rate depends on the
eccentricity. \label{rheology}This possibility seems to be precluded for solid,
homogeneous planets with an Andrade rheology \citep{ME13}. However, let us note
that planets with oceans may strongly depart from this predicted tidal
response. What is the frequency dependence of the tidal response of an
ocean-covered planet or of an "ocean planet," or even what is inventory of
water necessary to transition from one regime to the other remains poorly
constrained. Pseudosynchronous rotation thus remains a possibility to be
investigated.}
\item[$\bullet$] Venus, for example, tells us that thermal tides
in the atmosphere can force an asynchronous rotation \citep{GS69, ID78,
CL01,LWM15}. However, due to its 
scaling with orbital distance, this process seems to lose its efficiency around very low 
mass stars such as Proxima \citep{LWM15}.
\item[$\bullet$]Finally, if the orbit is still eccentric as might be the case here, 
trapping into a spin-orbit resonance, such as the 3:2 resonance of Mercury, becomes 
possible \citep{GP66}. 
\end{itemize}
The goal of this section is to quantify the likelihood of an asynchronous, resonant 
spin-orbit state. For sake of simplicity and concision, we will assume that the planet 
started with a rapid prograde spin. Because of the short spin evolution timescale, we will 
also assume that the obliquity has been damped early in the life of the system. We note, 
however, that trapping in Cassini states may be possible if the precession of the orbit 
is sufficient and tidal damping not too strong \citep{FJG07}. This possibility is left 
out for further investigations.

\subsubsection{Probability of capture in spin-orbit resonance}

For a long time, only few unrealistic parameterizations of the tidal dissipation inside 
rocky planets were available \citep{Dar1880,Lov09,Gol63}. At moderate eccentricities, 
models based on those parameterizations almost always predicted an equilibrium rotation 
rate -- where the tidal torque would vanish -- that was either synchronous or with a much 
slower rotation than the slowest spin-orbit resonance. As a result, tides would always 
tend to spin down a quickly rotating planet and persistence into a given resonance could 
only occur through \textit{trapping}. In this mechanism, the gravitational torque over a 
permanent, non-axisymetric deformation -- the triaxiality -- of the planet creates an 
effective ``potential well'' in which the planet can be trapped \citep{GP66}.

In this framework, consider a planet with a rotation angle $\theta$ and a mean anomaly $M$ 
(with the associated mean rotation rate, $\thetad$, and mean motion $\norb$) around an half 
integer resonance $\res$. Defining $\gam\equiv\theta-\res M$, the equation of the spin 
evolution averaged over an orbit is given by 
\balign{\C \ddot{\gam}= \Ttriax+ \Ttide,}
where \balign{\Ttriax\equiv-\frac{3}{2}\left(B-A\right) \Hpe \norb^2 \sin 2\gam}
is the torque due to the triaxiality $(B-A)/\C$ where $A, B,$ and $\C$ are the three 
principal moments of inertia of the planet (in increasing magnitude), $\Hpe$ is a Hansen 
coefficient that depends on the resonance and the eccentricity ($e$), and $\Ttide$ is the 
dissipative tidal torque. In their very elegant calculation, \citet{GP66} demonstrated that 
the probability of capture only depends on the ratio of the constant part of the tidal 
torque that acts to traverse the resonance to the linear one that needs to damp enough 
energy during the first resonance passage to trap the planet. This theory was further 
generalized by \citet{Mak12} who showed that this is actually between the odd and the even 
part of the torque (where $\Todd(-\gamd)=-\Todd(\gamd)$ and $\Teven(-\gamd)=\Teven(\gamd)$) 
that the separation needs to be done. With these notations, the capture probability simply 
writes
\balign{\label{captureprob}
\Pcapt=2/\left(1+\frac{\int_{-\pi/2}^{\pi/2} \Teven(\gamd)\d 
\gam}{\int_{-\pi/2}^{\pi/2}\Todd(\gamd)\d\gam}\right),}
where the integral should be performed over the separatrix between the librating (trapped) 
and circulating states given by
\balign{\gamd\equiv\width\cos \gam\equiv \norb\sqrt{3\frac{B-A}{\C}\Hpe}\cos \gam. 
\label{resonancewidth}}
For further reference, $\Delta$ will be called the width of the resonance, as this is the 
maximum absolute value that $\gamd$ can reach inside the resonance.

\eq{captureprob} is completely general and can readily be used with any torque. Hereafter, 
we will only use a tidal torque that is representative of the rheology of solid planets, 
i.e., the Andrade model generalized by \citet{Efr12}. Specifically, we will use the 
implementation in Eq.\,(10) of \citet{Mak12}. All model parameters are exactly the same as 
in this article (in particular, the Maxwell time is $\Tm=500$\,yr), except that we fix the 
Andrade time to be equal to the Maxwell time for simplicity. We note that
although this model and these parameter values fairly reproduce some features
from the tidal response of the Earth, it does not consistently account for the
effect of the oceans$^{\ref{rheology}}$. The numerical results for the capture
probability are shown where applicable for the 3:2 resonance in
\fig{fig:capture}.

\subsubsection{Is capture always possible?}

An interesting property of the solution above is that because it involves the ratio of two 
components of the torque, any overall multiplicative constant, i.e. the overall strength of 
tides, cancels out. At first sight, this seems to simplify greatly the survey of the whole 
parameter space because explicit dependencies on the stellar mass and orbital semi-major 
axis, among other parameters, disappear. The capture probability only depends on the 
eccentricity of the orbit, the triaxiality of the planet, and the ratio of the orbital 
period to the Maxwell time. This completely hides the fact that \textit{capture may be 
impossible even when the capture probability is not zero}. Indeed, as pointed out by 
\citet{GP66}, another condition must be met for trapping to occur: The maximum restoring 
torque due to triaxiality must overpower the maximum tidal torque inside the resonance. If 
not, even if the energy dissipation criterion is met, the tidal torque is just strong 
enough to pull the planet out of the potential well of the resonance. 

In our specific case, the maximum restoring torque and the maximum tidal torque trying to 
extract the planet from the resonance both occur at the lowest boundary of the resonance 
(when $\gamd=-\width$ and $\gam=-\pi/4$). This point is reached on the first swing of the 
planet inside the resonance, when it moves along a trajectory close to the separatrix. So, 
notwithstanding the value of $\Pcapt$, capture is impossible whenever 
\balign{\Ttide(\gamd=-\width) < -\frac{3}{2}\left(B-A\right) \Hpe \norb^2, }
both quantities being negative. This condition, hereafter referred to as condition (a), is 
verified below the curve with the same label in \fig{fig:capture}.

\subsubsection{Non-synchronous equilibrium rotation}

Contrary to simplified parametrizations of tides, the more realistic frequency dependence 
of the Andrade torque entails that synchronous rotation is not the only equilibrium 
rotation state. As illustrated by \fig{fig:equ-ecc}, depending on the eccentricity, the 
tidal torque can vanish for several rotation states, although only the ones near half 
integer resonances\footnote{As this process does not involve the same processes as the 
usual resonance capture, the ratio of equilibrium rotation rates to the mean motion are 
not exactly half integers.} are stable (see \citealt{Mak12} for details). For Proxima b's 
orbital period and with $\Tm=500$\,yr, the $\omega\approx3\norb/2$ rotation becomes stable 
for an eccentricity greater than 0.06--0.07, and the $\omega\approx2\norb$ above $e=0.16$.

\begin{figure} 
\subfigure{\includegraphics[scale=.7,trim = 0cm .cm 0.cm 0.cm, clip]{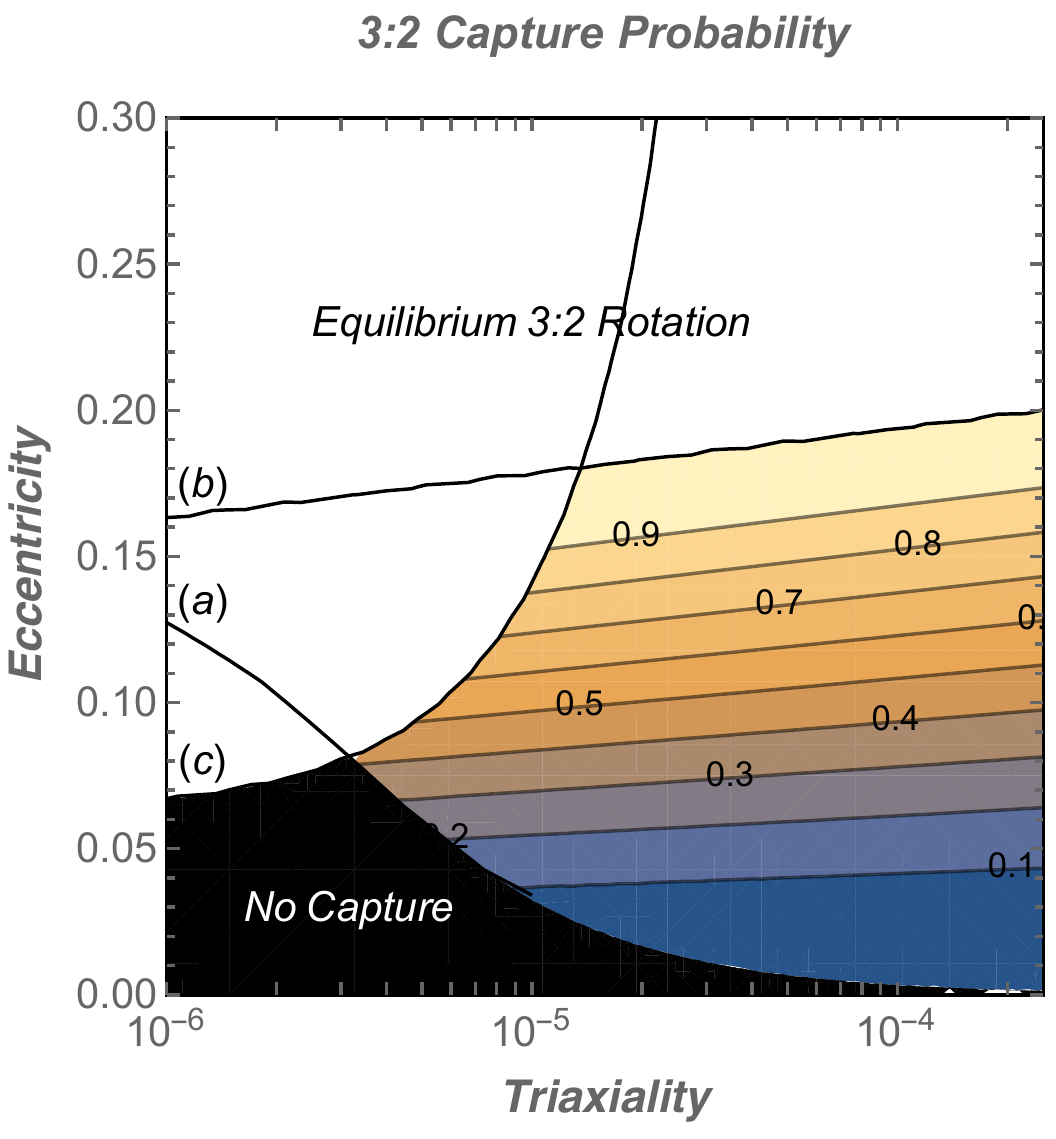}}
\caption{Probability of capture in the 3:2 resonance as a function of orbital 
eccentricity and triaxiality of the planet (numbered contours with color shadings). White 
regions depict areas of certain capture due to the tidal torque. Black regions show where 
the triaxial torque is too weak to enforce capture. Labeled curves are: (a) Tidal torque 
at the lower boundary of the separatrix is negative and greater in magnitude than the maximum
restoring torque, (b) Tidal torque at the lower boundary of the separatrix is positive, (c) 
Maximal Tidal torque inside the resonance is greater than the maximum triaxial torque. Being 
above (c) and/or (b) leads to certain capture. Below (c) and (a) capture is impossible.}
\label{fig:capture}
\end{figure}

\begin{figure} 
\subfigure{\includegraphics[scale=.7,trim = 0cm .cm 0.cm 0.cm, clip]{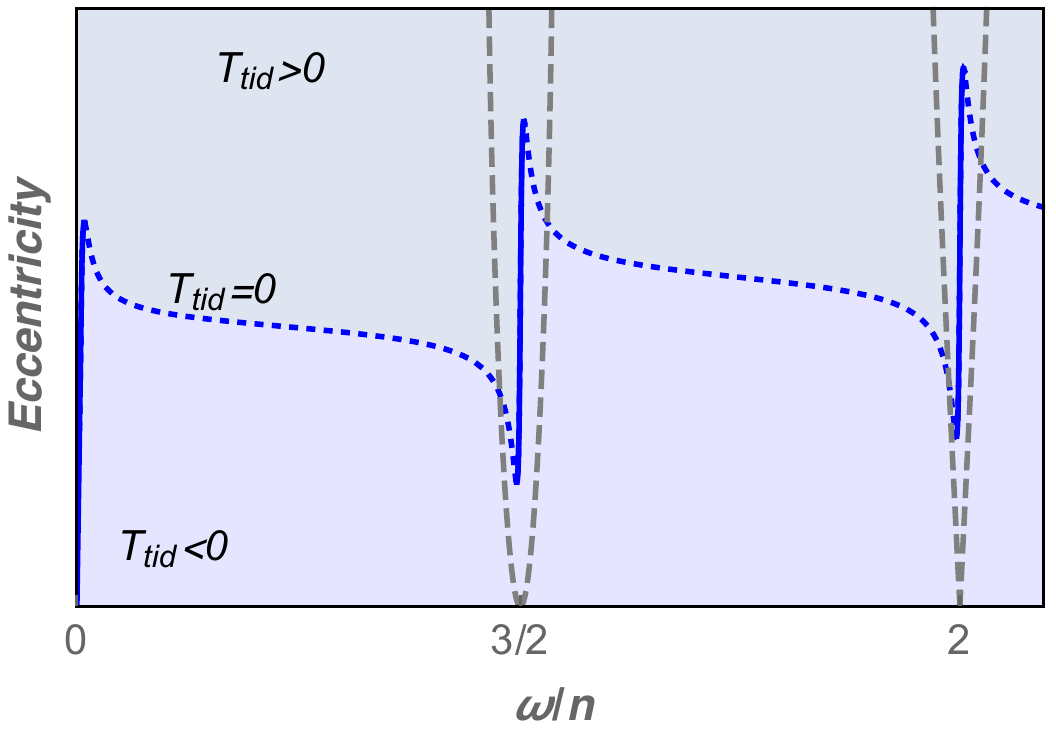}}
\caption{Sketch of the equilibrium eccentricity, whence the tidal torque vanishes, as 
a function of planetary rotation rate (blue curve). Solid portions of the curve near 
resonances depict stable equilibria, whereas dotted portions show unstable ones. Areas of 
positive torque are in gray, and of negative torque in blue. Dashed gray 
curves show the area covered by librations inside the resonances, i.e. the resonance 
width (see \eq{resonancewidth}). With realistic values for the triaxiality and the 
Maxwell time, both the kink around the resonances and the resonance width would be much 
narrower and difficult to see.}
 \label{fig:equ-ecc}
\end{figure}

In the absence of any triaxiality, the planet would always be stopped in the fastest 
stable equilibrium rotation state available for a given eccentricity. When triaxiality 
is finite, it entails libration around the resonance, inside the area delimited by dashed 
gray curves in \fig{fig:equ-ecc}. This can actually cause the planet to traverse the 
resonance. In such a case, the capture is probabilistic and its probability can be computed 
using \eq{captureprob}. There are however two conditions for which capture becomes certain:
\begin{itemize}
\item[(b)] If the tidal torque is positive at the lower boundary of the separatrix 
($\gamd=-\width$, i.e. along the dashed curve at the left of each resonance in 
\fig{fig:equ-ecc}), the planet is always brought back toward the equilibrium rotation.
\item[(c)] If anywhere in the resonance the tidal torque is positive and greater than the 
maximum triaxial torque, then the rotation rate can never decrease below that 
point.\footnote{If this situation occurs, the planet will never reach the bottom of the 
separatrix. Therefore, capture will ensue, even though the condition (a) is not met.} 
\end{itemize}
These two conditions are met above the black curves labeled (b) and (c) respectively in 
\fig{fig:capture}.

\subsubsection{Summary and implications for climate}

\fig{fig:capture} summarizes the chances of capture in the 3:2 resonance as a function of 
its triaxiality and eccentricity at the resonance crossing. The white area shows where 
capture is certain, and the black area, where capture is impossible. In the remaining part 
of the parameter space, capture probability is computed using \eq{captureprob}.
If capture does not occur, which may occur in the color shaded area and is
certain in the black one, the planet ends up in synchronous rotation. 

As expected, capture probability increases with eccentricity. We also recover
the fact that, at low eccentricities (here below $\sim$0.06), capture can only
occur if the triaxiality is sufficient to counteract the spin down due to tidal
friction. To put these numbers into context, let us note than triaxiality shows
variability from one planet to another, but seems to decrease with increasing
mass, from $\sim1.4\times10^{-4}$ for Mercury\footnote{Derived from the gravity
moments (in particular $C_{2,2}$) measured by \citet{SZP12}.} to
$\sim2\times10^{-5}$ for the Earth, and $\sim6\times10^{-6}$ for Venus
\citep{Yod95}. Being a little more massive, Proxima b's triaxiality is likely
to be smaller still. As a consequence, capture is rather unlikely for an
eccentricity below 0.06.

A slightly less intuitive result is that, at higher eccentricities, capture probability 
decreases when triaxiality increases. Again, this is due to the fact that, although a 3:2 
rotation rate might be an equilibrium configuration, triaxiality induced librations can 
help the planet get through the resonance. In this regime, the likely low triaxiality of 
the planet will probably trap the latter in the 3:2 asynchronous rotation resonance. 

In conclusion, let us recall that the final rotational state of an initially fast rotating 
planet will be the result of the encounter of several resonances. Moreover, both the 
eccentricity and the triaxiality of the planet could vary from one resonance encounter to 
the other. The final rotational state thus depends on the orbital history of the system. 
However, considering the range of eccentricities discussed above, it seems that resonances 
higher than 3:2 (or maybe 2:1) are rather unlikely. At eccentricities lower
than 0.06, the most probable state becomes the synchronous one. This highlights
the need for further constraints on the eccentricity of the planet,
its possible evolution, and the existence of additional planets.

As shown by \citet{Yang2013} and \citet{Kopparapu2016}, the inner edge of the HZ depends on 
the rotation rate of the planet. In particular, simulated atmospheres of synchronous 
planets with large amounts of water develop a massive convective updraft sustaining a 
high-albedo cloud deck in the substellar region. Based on these studies and the 
characteristics of Proxima, the runaway threshold is expected to be reached at $0.9$ and 
$1.5$~S$_{\oplus}$, for a non-synchronous and a synchronous planet, respectively 
($S_{\oplus}$ being the recent Solar flux at 1~AU). Assuming the planet's orbit did not 
evolve during the Pre-Main Sequence phase, it would have entered the HZ at $\sim$90~Myr if the 
rotation of the planet is synchronous, or at $\sim$200~Myr if the rotation of the planet is 
non-synchronous. Thus, before reaching the HZ, the planet could have spent 100--200~Myr 
in a region too hot for surface liquid water to exist. This can be compared to the Earth, 
which is thought to have spent a few Myr in runaway after the largest giant impact(s) 
\citep{Hamano2013}. During this stage all the water is in gaseous form in the atmosphere, 
and therefore it can photo-dissociate and the hydrogen atoms can escape.

\section{Water loss and volatile inventory}\label{waterloss}

Proxima~b has experienced a runaway phase that lasted up to $\sim$200~Myr, during 
which water is thought to have been able to escape. We discuss here the processes of water 
loss as well as the processes responsible for the erosion of the background atmosphere. 

\subsection{Modeling water loss}

In order to estimate the amount of water lost, we use the method of \citet{Bolmont2016} 
which is an improved energy-limited escape formalism. The energy-limited escape mechanism 
requires two types of spectral radiation: FUV (100--200~nm) to photo-dissociate water 
molecules and XUV (0.1--100~nm) to heat up the exosphere. We consider here that the planet 
is on a circular orbit at the end of the protoplanetary disk phase, its orbit thus remains 
constant throughout the evolution.
The mass loss is given by \citep{Lammer2003,Selsis2007a}:
\begin{equation}
\label{eqmassloss1}
\dot{m} = \epsilon \frac{F_{\rm XUV} \pi {R_{\rm p}}^3}{\mathcal{G} M_{rm p} 
(a/1 {\rm AU})^2},
\end{equation}
where $a$ is the planet's semi-major axis, $R_{\rm p}$ its radius and $M_{\rm p}$ its mass. 
$\epsilon$ is the fraction of the incoming energy that is transferred into gravitational 
energy through the mass loss. As in \citet{Bolmont2016}, we estimate $\epsilon$ using 1D 
radiation-hydrodynamic mass-loss simulations based on the calculations of 
\citet{OwenAlvarez2016}. For incoming XUV fluxes between 0.3 and 
200~erg~s$^{-1}$~cm$^{-2}$, the efficiency is higher than 0.1, but for incoming XUV 
fluxes higher than 200~erg s$^{-1}$ cm$^{-2}$, the efficiency decreases (down to 0.01 at 
$10^5$~erg s$^{-1}$ cm$^{-2}$, see Fig. 2 of \citealt{Bolmont2016}). $t_0$ is the 
initial time taken to be the time at which the protoplanetary disk dissipates. We 
consider that when the planet is embedded in the disk, it is protected and does not 
experience mass loss. We assume that protoplanetary disks around dwarfs such as Proxima 
dissipate after between $t_0=$3~Myr and 10~Myr 
\citep{Pascucci2009,Pfalzner2014,PecautMamajek2016}.

We consider here that the atmosphere is mainly composed of hydrogen and oxygen. From the 
mass loss given by Eq. (\ref{eqmassloss1}), we can compute the ratio of the escape flux 
of oxygen and hydrogen \citep{Hunten1987, LugerBarnes2015}. The ratio of the escape fluxes of 
hydrogen and oxygen in such hydrodynamic outflow is given by:
\begin{equation}
r_\textrm{F}=\frac{F_\textrm{O}}{F_\textrm{H}} = \frac{X_\textrm{O}}{X_\textrm{H}} \frac{m_\textrm{c}-m_\textrm{O}}{m_\textrm{c}-m_\textrm{H}}. \label{equ:mc}
\end{equation}
This ratio depends on the crossover mass $m_\textrm{c}$ given by:
\begin{equation}\label{escape2}
m_\textrm{c} = m_\textrm{H} + \frac{kT F_\textrm{H}}{bg X_\textrm{H}},
\end{equation}
where $T$ is the temperature in the exosphere, $g$ is the gravity of the planet and $b$ 
is a collision parameter between oxygen and hydrogen. In the oxygen and hydrogen mixture, 
we consider $X_\textrm{O}=1/3$, $X_\textrm{H}=2/3$, which corresponds to the proportion 
of dissociated water. 

\subsection{Water loss in the runaway phase}

To calculate the flux of hydrogen atoms, we need an estimation of the XUV luminosity of 
the star considered, as well as an estimation of the temperature $T$. We use the two different 
XUV luminosity prescriptions as in Sect. \ref{irradiation}, namely Proxima having had a 
saturation phase up to 3 Gyr and then a power-law decrease and another one with 
a constant value during its entire lifetime representing that saturation still lasts today
(see Fig. \ref{fig:XUVevol}). We adopt an exosphere temperature of 3000~K \citep[given by 
hydrodynamical simulations, e.g.][]{Bolmont2016}. In the following, we give the mass loss from 
the planet in units of Earth Ocean equivalent content of hydrogen (EO$_H$).

We calculated the hydrogen loss using three different methods: 
\begin{description}
\item[(1)] Assuming $r_{\rm F} = 0.5$, and calculating the mass loss as in 
\citet{Bolmont2016};
\item[(2)] Assuming $r_{\rm F} = f(F_{\rm XUV})$, and calculating the mass loss as in 
\citet{Bolmont2016};
\item[(3)] Computing the loss of hydrogen and oxygen atoms by integrating the expressions 
of $F_\textrm{O}$ and $F_\textrm{H}$ \citep[see the equations in][]{Bolmont2016}.
\end{description}

Using method (1) and (2) allows to bracket the hydrogen loss without doing the 
integration of method (3). Indeed, using $r_{\rm F} = 0.50$ allows to compute the best 
case scenario: the loss is stoichiometric, 1 atom of oxygen is lost every 2 atoms of 
hydrogen. However, using $r_{\rm F} = f(F_{\rm XUV})$ allows to compute the mass loss 
assuming an infinite initial water reservoir: whatever the loss of hydrogen and oxygen, 
the ratio $X_\textrm{O}/X_\textrm{H}$ remains the same and $r_{\rm F}$ only depends on 
$F_{\rm XUV}$.

\begin{figure}
\begin{center}
\includegraphics[width=\linewidth]{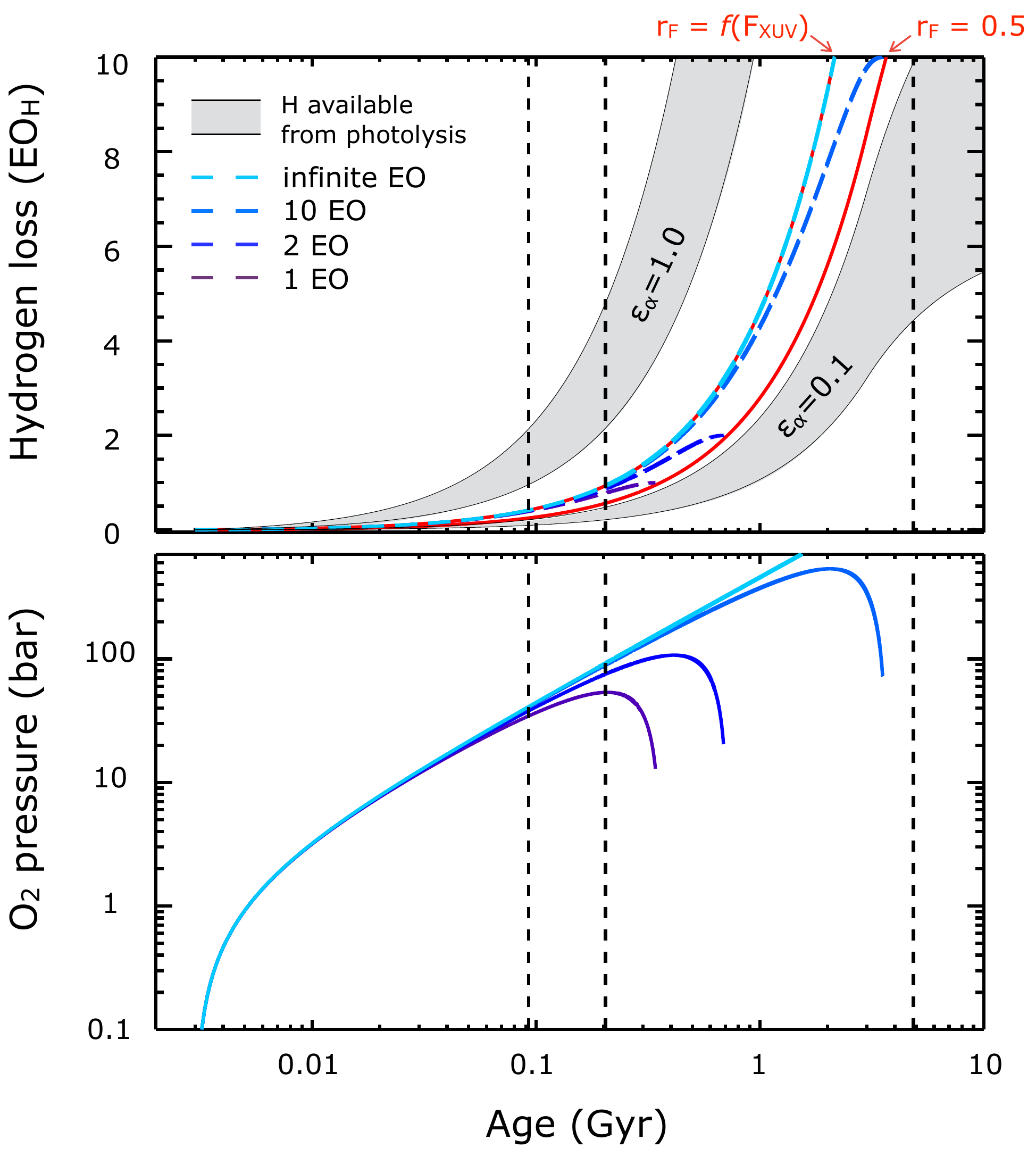}
\caption{Hydrogen loss and O$_2$ pressure for Proxima-b for an initial time of $t_0 = 
3$~Myr. Top panel: hydrogen loss computed with method (1) and (2) in full red lines 
and with method (3) in colored dashed lines. The grey areas correspond to the 
amount of hydrogen created by photolysis for two different efficiencies 
$\epsilon_\alpha = 1.0$ (all the incoming energy is used for photolysis) and 
$\epsilon_\alpha = 0.1$ (only 10\% of the incoming energy is used). The edges of the 
grey areas were calculated using two different assumptions on the wavelength range 
important for photolysis: the lower edge corresponds to the energy flux in the 
H Ly $\alpha$ band (Table \ref{tab:XUV}), the upper edge corresponds to the energy 
flux in a wider band: 10--170~nm. The vertical lines represent the time at which 
the planet reaches the 1.5~S$_{\oplus}$ HZ inner edge, the 0.9~S$_{\oplus}$ 
HZ inner edge and the age of the system (4.8~Gyr). Bottom panel: O$_2$ pressure
building up in the atmosphere computed with method (3).}
\label{Hloss_PO2_3_10Myr}
\end{center}
\end{figure}

\begin{table*}
\centering
\caption{Lost hydrogen (in EO$_H$), lost oxygen (in bar) and atmospheric build-up O$_2$ 
pressure when Proxima b reaches the HZ and at the age of the system. The two values given 
for each column correspond to the uncertainty coming from the different initial water 
reservoir (1 ocean to $\infty$ oceans) and the initial time ($t_0 = 10$~Myr and 
$t_0 = 3$~Myr).}
\vspace{0.1cm}
\begin{tabular}{|c||c|c|c||c|c|c||c|c|c|}
\hline
 \multicolumn{1}{|c||}{}  	& \multicolumn{3}{c||}{H loss (EO$_H$)}  & \multicolumn{3}{c||}{O$_2$ loss (bar)} & \multicolumn{3}{c|}{build-up O$_2$ pressure (bar)} \\
\multicolumn{1}{|c||}{}	& $T_{\rm HZ}$      & $T_{\rm HZ}$	    & 4.8~Gyr   & $T_{\rm HZ}$      & $T_{\rm HZ}$ 	    & 4.8~Gyr	& $T_{\rm HZ}$ 	    & $T_{\rm HZ}$      & 4.8~Gyr \\
\multicolumn{1}{|c||}{}	& (1.5~S$_{\oplus}$)& (0.9~S$_{\oplus}$)&           & (1.5~S$_{\oplus}$)& (0.9~S$_{\oplus}$)& 	        & (1.5~S$_{\oplus}$)& (0.9~S$_{\oplus}$)& \\
\hline
$L_{\rm XUV}$ evol  & 0.36--0.42 	& 0.76--0.94 	& $<$21	& 47--51    & 110--118 	& 207--2385		& 32--41	& 53--92	& $<$2224	\\
$L_{\rm XUV}$ cst   & 0.25--0.29 	& 0.55--0.64 	& $<$16 & 21--22    & 48--52 	& 176--1193		& 35--41	& 71--92 	& $<$2224	\\
\hline
\end{tabular} 
\label{waterlost1}
\end{table*}

With this method we can compute the hydrogen loss from Proxima b. Fig. 
\ref{Hloss_PO2_3_10Myr} shows the evolution of the hydrogen loss with time for 
an initial time of protoplanetary disk dispersion of 3~Myr assuming different 
initial water reservoirs and with the different methods. Table \ref{waterlost1} 
summarizes the results for the two different XUV prescriptions: $F_{\rm XUV} = {\rm cst}$ 
and $F_{\rm XUV} = {\rm evol}$ (as given in Sect. \ref{irradiation}). We used for these 
calculations the minimum mass of Proxima b (1.3~M$_{\oplus}$). If a mass corresponding 
to an inclination of 60$^\circ$ is adopted ($\approx$1.6~M$_{\oplus}$; the most probable 
one) the resulting losses are slightly higher but by no more than about 10\%, which is 
negligible given the uncertainties in other parameters.

We find that for the evolving XUV luminosity the water loss from the planet is below 
0.42~EO$_H$ at $T_{\rm HZ}$ (1.5~S$_{\oplus}$) and below $\sim$1~EO$_H$ at $T_{\rm HZ}$ 
(0.9~S$_{\oplus}$). The loss of hydrogen does not significantly change when considering 
different initial time of protoplanetary disk dispersion (3 or 10~Myr here). The 
calculations thus suggest that the planet does not lose a very high amount of water 
during the runaway phase. 

Fig. \ref{Hloss_PO2_3_10Myr} also shows the hydrogen produced by photo-dissociation (grey 
areas in top panel). 
If all the incoming FUV photons do photolyse H$_2$O molecules with $\epsilon_\alpha = 1$ 
(100\% efficiency) and if all the resulting hydrogen atoms then remain available for the 
escape process then photolysis is not limiting the loss process. However, when considering 
a smaller efficiency ($\epsilon_\alpha = 0.1$), we can see that photo-dissociation is the 
limiting process, indeed hydrogen is being produced at a slower rate than the escape rate.
We find that for $\epsilon_\alpha < 0.2$, photo-dissociation becomes the limiting process 
for the stronger hypothesis on the FUV incoming flux (i.e., when we consider all that is 
emitted between 10 and 170~nm). 

Because oxygen is lost at a much slower rate than hydrogen (but is still lost quite fast, 
see Table \ref{waterlost1}: up to 50~bar of O$_2$ is lost before the planet reaches the 
1.5~$S_\oplus$ HZ), the loss of hydrogen results in a build-up of oxygen in the atmosphere. 
We calculated the amount of remaining oxygen (that may be removed from the atmosphere by 
chemical reactions with surface minerals and recycling of the crust) as a potential O$_2$ 
pressure in the atmosphere. 
If the inner edge of the HZ is defined by $S_{\rm p}=1.5$~S$_{\oplus}$, the resulting 
O$_2$ pressure is of the order of 30 to 43~bar. Assuming the planet initially has a water 
content equal to 1 Earth ocean, the O$_2$ pressure starts decreasing as the hydrogen 
becomes scarce and oxygen becomes the only species to escape. This explains the wide 
range of O$_2$ pressures at $T_{\rm HZ}$ (0.9~S$_{\oplus}$): from 55~bar to almost 
100~bar. Of course, had we assumed an atmosphere with more species, we expect 
that the species reaching sufficiently high up in the atmosphere to escape as well.
The presence of a background atmosphere would probably slow down the escape of hydrogen, 
which means our calculations are an upper value on the hydrogen loss.

\subsection{Nitrogen loss associated with the hydrogen escaping flow} \label{sub:nitrogen}

Just as oxygen atoms, nitrogen atoms can be dragged away by collision with the outflow of 
hydrogen. Compared with other volatile elements, and assuming a carbonaceous chondrite 
origin, nitrogen is depleted on Earth by one order of magnitude. There is about as much 
nitrogen in the atmosphere and in the mantle of the Earth \citep{marty12}, the missing 
part being possibly trapped into the core \citep{Roskosz2013} since the differentiation of 
the planet. Unless Proxima~b accreted a much larger initial nitrogen amount and/or suffered 
less nitrogen segregation into its core, the atmospheric escape of nitrogen -- an element 
essential to life as we know it -- represents a major threat to its habitability.

Oxygen loss rates computed in the previous section show that hydrodynamic escape of 
hydrogen can potentially exhaust several bars of nitrogen reservoir in a few Myrs.
By the time the planet reaches the HZ defined by $1.5~S_\oplus$, the equivalent of $50$~bar 
of O$_2$ can be lost (see Table \ref{waterlost1}). A similar quantity of N$_2$ could be lost, 
which represents more than 30 times the reservoir of nitrogen in the Earth's atmosphere and 
mantle. Two effects should however protect the nitrogen reservoir. First, during the 
runaway phase, the atmospheric content is in equilibrium with the magma ocean and a 
fraction of the nitrogen reservoir is therefore in solution in the interior. This partition 
is not sufficient in itself to protect nitrogen from escaping. Indeed, equilibrium between 
the silicates and the atmosphere implies that mantle outgassing compensates for the loss 
to space, which therefore exhausts both reservoirs. Vertical transport and fractionation 
within the mantle may, however, deplete the upper mantle that exchanges with the atmosphere 
and bury nitrogen at depth by the same process that may have enriched the core of the Earth 
in nitrogen. If part of this nitrogen is lost into the core another part is outgassed when 
the mantle solidify after water condensation at the end of the runaway 
\citep{2008E&PSL.271..181E}. As the solubility of nitrogen into silicates depends strongly 
on the redox state of the mantle and its water content \citep{Roskosz2013} it is difficult 
to draw quantitative conclusions but current geochemistry tells us that part of the 
nitrogen should indeed be released only once the planet enters the HZ. 

Another strong limitation of nitrogen escape is the photolysis of N$_2$. Due to its triple 
bond, N$_2$ is photodissociated only in the 40--100~nm wavelength range. But there appear
to be sufficient photons (even with the current stellar emission) to photolyze 1 bar of 
N$_2$ in a few Myrs. Assuming that no other species absorbs and that atoms released by N$_2$ 
photolysis do not recombine, this is about enough to feed the computed loss rates. However, 
losing N atoms through the hydrogen outflows implies that the atmosphere is dominated by 
H$_2$O and that H$_2$O is efficiently photolysed to produce H atoms at least as fast as 
they are lost. Water vapor absorbs at the wavelengths photolyzing N$_2$ (40--100~nm) with a 
similar cross-section ($\sim$10$^{-17}$~cm$^2$) and photons are thus absorbed by a column 
of $10^{17}$~cm$^{-2}$ of either of the two species. If such column is found above the 
N$_2$--H$_2$O homopause then H$_2$O, which has a 50\% larger scale height, absorbs most of the 
incoming radiation even if both species have similar mixing ratios at the homopause. To 
determine the total concentration at the homopause we must assume a value for the eddy 
mixing coefficient $K_{ZZ}$. Near the homopause of the terrestrial planets of the solar 
system $K_{ZZ}$ spans values from $3\times10^{5}$~cm$^{2}$~s$^{-1}$ on Earth to 
$10^{8}$~cm$^{2}$~s$^{-1}$ on Mars \citep{Atreya1999}. Using a H$_2$O--N$2$ diffusion 
parameter of $6\times10^{18}$~cm$^{-1}$~s$^{-1}$ \citep{Chamberlain_Hunten1987}  we can 
determine the homopause concentration $n_h$ as well as the column of H$_2$O and N$_2$ 
assuming a temperature at the homopause (we take 300~K) and a mixing ratios $X_{h2o}$, 
$X_{n2}$ for the two species below the homopause. We find a H$_2$O column density between 
$X_{h2o}\times8.5\times10^{16}$  and $X_{h2o}\times2.8\times10^{19}$~cm$^{-2}$ 
and a N$_2$ column density between $X_{n2}\times5.4\times10^{16}$ and 
$X_{n2}\times1.8\times10^{19}$~cm$^{-2}$. For low values of $K_{ZZ}$, photons are absorbed
above the homopause by H$_2$O even for significant N$_2$ mixing ratios. (Note that in this 
case, we verified that the altitude of absorption always remains negligible compared with 
the radius of the planet, even for hot expanded thermospheres, so that the UV absorption 
section of the planet is not enhanced by the expansion of the upper atmosphere). For high 
values of $K_{ZZ}$, photolysis of H$_2$O and N$_2$ occurs within the well mixed atmosphere 
and photolyzing photons will be absorbed in majority by the most abundant species. In 
models of Earth early atmosphere in equilibrium with a magma ocean, the water vapor 
pressure is estimated at several hundreds of bars. Assuming the present silicate/atmosphere 
partition for nitrogen (which, as said previously could have been in fact dominated by the 
mantle reservoir) yields a mixing ratio of less than $0.01$. Accounting also for the 
recombination of nitrogen atoms, the integrated loss rate of nitrogen falls well below 
a bar of N$_2$ in 100~Myr.

Carbon atoms can also be dragged by hydrogen but are expected to be protected by the same 
processes as nitrogen. Details would be quantitatively different because carbon dioxide -- 
assumed to be the carrier -- absorbs over a different wavelength range (still overlapped by 
that of H$_2$O) and has mass 60\% larger than N$_2$. Based on Venus and Earth carbon 
inventories (equivalent to more than 90~bars of CO$_2$ for Venus and 150~bars for the 
Earth), we know that the carbon reservoir can exceed by far the nitrogen content and the 
amount that can be exhausted by hydrogen drag alone.

\subsection{Water loss while in the HZ}

Let us consider a scenario where Proxima~b enters the HZ with enough water to 
condense into an ocean but with only a tenuous background atmosphere of 10~mbar or less. 
\citet{Turbet16} show that this configuration is still compatible with surface habitability 
for large amounts of water preventing its trapping at the poles (3:2 spin-orbit resonance) 
or on the night side (synchronous rotation). But they also show that the tropospheric cold 
trap of water is no longer efficient in this case resulting in a H$_2$O-rich upper 
atmosphere. This is due to the fact that, in the substellar region, the vapor pressure 
at the surface exceeds the pressure of the background gas and the decrease vapor pressure 
with altitude (following the temperature) is not steeper than the decrease of the 
background pressure. The upper atmosphere is therefore fed with H$_2$O through substellar 
warm vapor-rich columns.

In the absence of a dense enough background atmosphere, the escape of oxygen and 
hydrogen could therefore continue on much longer timescales without being limited by the 
diffusion of water vapor. If we integrate the water loss with the same assumptions used for 
the runaway phase (except for the evolution of the XUV flux, which is accounted for) over 
the whole history of the planet (4.8~Gyr) we find that the planet loses as much as 
$\sim$21~EO$_H$ by the age of the star (see Table \ref{waterlost1}). Considering the 
constant XUV luminosity assumption, the loss decreases to $\sim$16~EO$_H$ at the current age 
of the system. As a consequence, a situation with Proxima~b in the HZ with a global ocean 
but a surface pressure lower than a few tens of mbars is likely to result in the rapid 
($<$1~Gyr) total loss of water and volatiles. Indeed, it would be difficult to imagine 
a reservoir of water larger than 16~EO$_H$ with such a depletion of all other 
atmospheric species.

\subsection{Other mechanisms for the loss of heavy elements}

Heavy elements like carbon,oxygen and nitrogen escape at much slower rate than hydrogen 
under similar conditions but, at the exception of O/H$_2$O, the transport of their main 
carrier (N$_2$, CO$_2$) toward the upper atmosphere is not limited by condensation like 
water. Therefore their loss proceeds through the whole history of the planet and not only 
when they are carried away by a hydrogen outflow as discussed in section \ref{sub:nitrogen}. 
In the absence of hydrogen, however, a significant erosion of the CO$_2$ or N$_2$ reservoir 
of an Earth-like planet would require XUV fluxes much higher than that experienced by 
Proxima~b. \citet{Tian2009} estimated for instance that a flux of 150~XUV$_\oplus$ would 
erode less than 0.5\% of the Earth carbon reservoir from a 5.9~M$_\oplus$ planet during 
1~Gyr. Although this calculation was done for a large terrestrial planet, with twice the 
gravity of a $1.3$~M$_\oplus$ rocky planet, other calculations by \citet{Tian2009} show 
that the loss becomes weakly gravity-dependent at very high XUV fluxes and that only 
fluxes higher than 1000~XUV$_\oplus$ could erode more than half the Earth carbon reservoir. 
Therefore, if an initial reservoir of heavy volatile elements comparable to that of Earth 
or Venus was present on Proxima~b at the end of its accretion, the majority of it should 
not have been lost through thermal escape alone.

The loss process that represents the main threat for the survival of a dense atmosphere on 
Proxima~b is the joint effect of EUV-driven atmospheric thermal expansion and non-thermal 
losses, mainly ion pick-up by particles from the stellar wind \citep{Lammer2011} and CMEs  
\citep[coronal mass ejections][]{Khodachenko2007,Lammer2007}. When the Sun has a low 
activity, the Earth exobase (the limit above which collisions become negligible) is found 
at $\sim$500~km far below the magnetopause at $\sim$15 planetary radii ($96\,000$~km). 
Inside the magnetopause the plasma dynamics are controlled by Earth magnetic field while 
outside they are driven by the solar wind. During high activity phases of the Sun, the 
EUV flux received by the Earth increases by a factor of two. As a consequence the exobase 
rises to $\sim$1000~km while the magnetopause moves inward, down to $\sim$7~R$_\oplus$ 
($45\,000$~km)  due to an increased solar wind pressure. Under higher EUV fluxes and stellar 
wind pressure the magnetopause can be compressed below the exobase and neutral species 
ionized above the magnetopause by either charged particle impacts or XUV irradiation are 
efficiently picked-up through open magnetic field lines. 

The exposure of Proxima~b to stellar wind could be moderate to severe, with a particle 
flux 4--80 times higher than that received by the Earth (see Sect. 
\ref{irradiation}). Non-thermal losses depend therefore more on the atmospheric 
expansion driven by XUV heating relative to the magnetopause distance maintained by 
the magnetic field of the planet. We found that the magnetopause radius of Proxima~b could 
be between 1.3 to 7 planetary radii depending on the values of the intrinsic magnetic 
moment, on the stellar wind pressure and on whether the stellar magnetic field is 
dipole-dominated or multipolar (see Sect.~\ref{magnetopause}). \citet{Tian2008_1} 
found that Earth exobase would rise to 4.8 and 12.7 planetary radii for 10 and 20 
EUV$_\oplus$, respectively. This result depends, however, strongly on the details 
of the atmospheric composition and in particular on infrared coolants. \citet{Tian2009} 
calculated that a pure CO$_2$ atmosphere requires much higher EUV fluxes to expand to 
similar distances (for instance, 1000~EUV$_\oplus$ for an exobase at 5 planetary radii). 
We can see that, within the possible range of magnetopause and exobase distances, the 
magnetosphere could penetrate or could not the atmosphere. A similar question exists 
for the early Earth: Using estimates for the evolution of Solar EUV/wind 
\citet{Lichtenegger2010} calculated that the magnetopause could have been as 
close as 2--4~R$_\oplus$ during the first Gyr of Earth evolution (when the Solar 
wind at Earth distance is assumed to have been much stronger than it is now and 
was at Proxima~b). Whether this distance provided a protection against non-thermal 
escape depends on the atmospheric composition. \citet{Lichtenegger2010} claimed 
that a CO$_2$-rich atmosphere could have withstood erosion while a N$_2$-rich 
atmosphere with a low CO$_2$ content (few percent or less) would have resulted 
in the loss of Earth's nitrogen reservoir. Estimating the loss by this process 
would require to explore the response of different types of atmospheres to EUV 
heating for different planetary magnetic moments. But it means also that the Earth 
kept an atmosphere despite a severely compressed magnetosphere and the expanded 
atmosphere predicted by models. 

In addition to the stellar wind itself, particles emitted by CMEs impact the planet 
sporadically, potentially compressing the magnetosphere within the atmosphere and eroding 
it by ion pick-up \citep{Khodachenko2007,Lammer2007}. These authors estimated that CMEs 
could prevent terrestrial planets in the habitable zone of low-mass stars like Proxima from
having a dense atmosphere. This conclusion is based on the combination of enhanced stellar 
wind pressures, high XUV heating and inefficient magnetic protection due to the slow 
rotation of tidally-evolved planets and the weak resulting dynamo. They estimated the 
loss rate from a CO$_2$-rich atmosphere as a function of the XUV flux and for different 
magnetopause distances. If we consider the present XUV flux at Proxima~b (60~XUV$_\oplus$) 
and a magnetopause at 2~R$_\oplus$, less than 1~bar of atmosphere would be lost per Gyr. 
A less magnetized planet with a magnetopause at 1.5~R$_\oplus$ would lose a few bars and 
a non-magnetized planet would be stripped off its atmosphere at a rate of more than 
100~bars per Gyr. If we assume that the XUV flux was higher in the past, up to 
150~XUV$_\oplus$, then even a magnetopause at 2~R$_\oplus$ could not prevent the planet to 
lose hundreds of bars/Gyr.

We should note, however, that these pioneer attempts to calculate non-thermal losses still 
need to be improved in several ways. First, they lack a determination of an energy balance that 
would allow to determine what the maximum mass loss rate is and what limits the loss in a 
particular case. For instance, in the case of infinitely efficient thermal loss above the 
magnetopause, the mass loss would become limited by the expansion of the atmosphere above 
the magnetopause and would thus be related with the thermal hydrodynamic expansion and the 
XUV energy-limited loss rate, unless particles contribute also to this expansion. 

In addition, in the absence of an intrinsic magnetic moment, ionospheric motions induce a 
magnetic field. On Venus for instance, the stellar wind does not penetrate the atmosphere 
that remains protected by this induced field \citep{Zhang2007}. 
It should be noted that the stellar magnetic activity that results in non-thermal escape 
also contributes to shield the atmosphere. For this reason, the magnetopause is more 
extended during active phases of the Sun than at Solar minimum \citep{Zhang2007}. 
Theoretical studies concluded that a magnetic field of 0.03--0.3~G is induced in the 
ionosphere even when Earth's magnetic moment disappears during magnetic reversals 
\citep{Birk2004}. Atmospheres exposed to a strong ionizing environment as well as a 
strong stellar wind or CMEs are expected to develop induced magnetic moments much stronger 
than those found in the solar system. In particular, the formation of so-called 
circumplanetary plasma-disks is now studied as a natural consequence of such exposition for 
close-in planets \citep{Khodachenko2012}. Previous estimates of non-thermal losses will 
have to be revisited to account for this additional and, potentially efficient, 
protection.

Thermal and non-thermal loss rates that have been estimated in previous studies are usually 
given as a function of EUV$_\oplus$ or XUV$_\oplus$. Soft X-rays and EUV have, however, 
very different absorption cross-sections and interact differently with atmospheric atoms 
and molecules. The fraction of incoming energy that actually powers the atmospheric 
expansion strongly depends on the altitude of absorption. Proxima b receives 250, 30 and 15 
times more flux than the Earth in of X-rays, EUV and Ly$\alpha$, respectively, and 60 times 
more in the overall XUV range. Results from the literature are therefore not directly 
applicable to Proxima b when they are given only in terms of EUV$_\oplus$ or XUV$_\oplus$, 
especially when they are based on solar or early-Sun spectra.

\subsection{Lessons from planet Earth}

As highlighted above, estimates of volatile loss to space are subject to large and even 
unknown uncertainties. First, they rely on complex models that were never confronted to 
actual observations of massive escape. None of the available models include all the mechanisms
controlling the loss rate, e.g., the photochemistry of the upper atmosphere and its detailed 
interaction with the wavelength-dependent stellar emission, non-LTE cooling processes, and an 
accurate description of the outflow beyond the exobase where hydrodynamics no longer apply. 
Some key data are not known, like the intrinsic planetary magnetic moment, now and in the past, 
the detailed evolution of the atmospheric composition, of the high-energy spectrum and of 
stellar wind properties. Maybe more importantly, the erosion of the volatile inventory 
critically depends on the time-dependent distribution of the volatiles between the atmosphere 
and the different layers of the interior, which is currently beyond predictive modeling.

Considering this situation let us discuss another approach by comparing Earth and Proxima b 
early history. It took about 100~Myr to build the Earth by colliding tens of Moon- to 
Mars-sized protoplanets \citep[see][]{morby12,raymond14}. Each of these giant collisions 
resulted in a runaway and magma-ocean phase that lasted 1 to a few Myr \citep{Hamano2013}. 
Therefore, protoplanets that eventually became the Earth spent a large fraction of these 
100 Myr in such runaway phase while exposed to the early Sun high-energy radiation. At that 
time, the Sun was likely to be in its saturated phase, characterized by a maximum 
L$_{\rm X}/$L$_{\rm bol}$ ratio. The XUV irradiation and stellar wind on the proto-Earth 
was therefore comparable, and possibly higher, than that of Proxima b. Proxima~b spent 
100--200 Myr in runaway before entering the HZ, which is longer than the runaway phases 
experienced by the proto-Earth by a small factor only ($<$10). Models predict that early 
Earth suffered massive volatile losses: hydrodynamic escape of hydrogen dragging away 
heavier species and non thermal losses under strong stellar wind exposure and CMEs 
\citep{Lammer2012}. Nonetheless, no clear imprint of these losses is found in the present 
volatile inventory.

The main source of Earth volatiles is believed to be carbonaceous chondrite (CC) material 
that, on average, shares with our planet a similar volatile pattern both in terms of relative 
abundances and isotopic signatures, in particular with identical D/H and $^{14}$N/$^{15}$N 
ratios \citep{Marty2012}. No mixture of other known sources of volatiles (for instance 
cometary and solar) can reproduce these patterns. Nitrogen found in the silicate Earth and 
the atmosphere is depleted by one order of magnitude compared with a CC composition, but 
there is mounting evidence that this missing nitrogen was segregated into the core 
\citep{Roskosz2013}. Some lighter hydrogen leaks from the deep mantle but is attributed to 
a solar nebula origin as confirmed by associated noble gases \citep{Marty2012}. From 
geochemical evidence alone there is no indication of a significant erosion of the Earth 
volatile inventory and one possible conclusion is that the volatile loss from the Earth 
was very limited. The only depleted and (slightly) fractionated species is Xenon. This 
could have been caused by atmospheric escape but considering that lighter species like 
Krypton are not depleted nor fractionated, it requires some complex scenario coupling 
preferential exchanges with mantle \citep[e.g. ][]{Tolstikhin2014} and/or non-thermal 
escape processes affecting only Xenon. For instance Xenon could be the only noble gas 
that can escape as an ion \citep{Zahnle_LPI_2015}. This scenario should, in addition, 
apply to the completely different history of Mars where Xenon presents the same puzzling 
properties. Until a robust explanation is found for the missing Xenon, its case cannot really 
provide any constraints on atmospheric escape.

Overall, escape processes produced negligible fractionation and preserved CC volatile 
patterns. Integrating these losses -- that did not alter Earth's volatile inventory in 
a measurable way -- over a runaway period about ten times longer on Proxima b does not 
appear as an overwhelming threat to habitability.

\section{Conclusions}

This paper presents a first assessment of the potential habitability of the 
recently announced Earth-like planet Proxima b. Different aspects including the initial
water inventory, history of the stellar XUV and particle emissions, orbital evolution, 
volatile losses, including water and the background atmosphere before and after entering 
the HZ, have been evaluated. The main summary and conclusions from our study are:
\begin{itemize}
\item[-] We cannot currently constrain Proxima b's initial water content. The planet's water 
budget was determined by the planet's feeding zone during its formation. Given its much 
smaller orbital radius it must have accreted much faster than Earth. We can envision a 
range of plausible formation scenarios that cover a broad range of volatile contents, 
from nearly dry planets to Earth-like water contents to waterworlds.  
\item[-] Using observations of Proxima, we estimate its high-energy emission in several 
wavelength ranges (from 0.6 to 170~nm). The total XUV irradiation 
that Proxima b receives today is of 307~erg~s$^{-1}$~cm$^{-2}$ between 0.6 and 118 nm, 
and 147~erg~s$^{-1}$~cm$^{-2}$ in the far-UV between 118 nm and 170 nm. Note that, on 
a short timescale, the high-energy flux is composed of quiescent emission plus
short-term flare events covering a variety of energy levels. The value we provide is a 
time average over a timescale of years that can be used as energy input to the 
planetary atmosphere.
\item[-] Proxima b receives 60 times more XUV flux than the current Earth, and 10 times
more far-UV flux. Note that the high-energy emission spectrum of Proxima is significantly 
harder than that of the Sun today, i.e., the relative contribution of X-rays compared to 
EUV is much higher. 
\item[-] We also give a prescription for the XUV-flux evolution of Proxima on nuclear 
timescales (from the time of stellar formation until the present age -- 4.8 Gyr). We 
adopt two different scenarios that should represent the extreme cases bracketing the 
real behaviour. One is an XUV irradiation on Proxima b $\sim$150 times stronger that the 
Earth today during the first 3 Gyr of its lifetime followed by a decay in the form of a 
power-law relationship. The other one is a constant flux in time, namely $\sim$60 times 
stronger than the Earth today. The total integrated XUV dose of Proxima b should be in
the range 7--16 times what Earth has received.
\item[-] Indirect observations yield an upper limit of the mass loss rate of Proxima of 
0.2~$\dot{\rm M}_\odot$. Assuming different geometries, we estimate that Proxima b 
is receiving a particle flux that should be at most a factor 4--80 higher than today's 
Earth value and roughly constant for the entire lifetime of the planet.
\item[-] We carry out a tidal analysis of the system by taking into account the host star 
Proxima. On the one hand, the tidal evolution of Proxima induced by the stellar tide is  
negligible even when assuming a very high dissipation factor for the star. Therefore, the 
history of the star, and in particular the history of its rotation period, does not have an 
influence on the orbit of Proxima b. On the other hand, we find that the tide raised by 
Proxima in the planet leads to a very small decrease in eccentricity and semi-major axis. 
We therefore find that, at the age of the system, Proxima b may have retained most of its 
initial eccentricity (which could be as high as 0.1).
\item[-] We determine the possible spin states of Proxima b at the age of the system. Due to 
the possible remnant eccentricity, the planet could have been captured in a 3:2 spin-orbit 
resonance depending on its level of triaxiality. If the orbit is circular or near circular, 
the planet could be synchronized. For realistic triaxialities, the transition occurs around 
$e=0.06$.
\item[-] Proxima b experiences two distinct major phases in its evolution: \emph{(1)} During 
the first few million years after the protoplanetary disk dispersion the planet is too hot 
for surface liquid water to exist; and \emph{(2)} After the first few hundred million years the 
planet enters the HZ. During these two phases Proxima b experiences atmospheric loss.
\item[-] We find that during the initial runaway phase the planet loses less than 1~EO$_H$.
We cannot provide strong constraints on the survival of the background atmosphere during 
the runaway phase: with the harshest constraints, up to $\sim$100 bar of N$_2$ could be 
lost. Under these circumstances, once in the HZ most water would be in the high atmosphere 
and susceptible to escape. In this case, Proxima b could have lost 16--21~EO$_H$ by the age 
of the system. It is also possible the planet has been able to keep its background 
atmosphere over the 4.8~Gyr of evolution. In this case the total water loss would far lower.
\item[-] During the runaway phase the erosion of a significant water reservoir would have 
left large amounts of residual oxygen, possibly in the form of atmospheric O$_2$, of up 
to 100~bar. This may have two important astrobiological consequences. First, the atmosphere, 
surface, ocean, and crust could have been strongly oxidized at the time the planet entered 
the HZ, which could prevent prebiotic chemical processes important for the origin of life. 
Second, a search for biosignatures must account for a possible abiotic build-up of O$_2$ 
and the consequent formation of an ozone layer.
\item[-] Due to XUV-driven atmospheric expansion and compression of the magnetopause by 
stellar wind, the planet could have lost a significant fraction of its volatile content 
by non-thermal losses. But quantifying the erosion depends on many unknown factors 
like the intrinsic and induced magnetic field, the actual stellar wind pressure on the 
planet, or the atmospheric composition. Further constraints are needed to assess whether
Proxima b has suffered intense atmosphere losses via these mechanisms or whether it has 
been able to keep its volatile inventory during its lifetime. 
\end{itemize}

Our results highlight the difficulty of assessing the habitability of a planet with 
an environment so different to Earth's and with only a rough estimate of its past 
irradiation and orbital history. Some of the scenarios we consider lead to a massive loss of 
volatiles, including water and the background atmosphere, and thus to a desiccated, 
inhospitable planet. But when other, equally plausible, constraints are used, the current 
state of Proxima b turns out to be one in which a dense atmosphere is still present and 
that only a modest amount of water (perhaps less than 1 Earth Ocean) has been lost to space.
Of course, it all depends on the initial volatile content and on the efficiency of 
processes that counteract themselves, namely in protecting or enhancing the erosion of the 
atmosphere subject to high-energy irradiation and particle wind. Although we have been 
able to present new, quantitative information on the high-energy and particle environment
of Proxima b, significant uncertainties in the initial state and subsequent evolution 
remain. The case of Venus and the Earth in the Solar System also provide interesting 
comparisons. In spite of the strong solar emissions in the past, they both have managed 
to keep dense atmospheres and no conclusive evidence of massive volatile loss has 
been found. This is even so in the case of Venus, which has no protection from an
intrinsic magnetic field. 
 
Much research is still due, but the main general conclusion from our study is that 
Proxima b could have liquid water on its surface today and thus can be considered a 
viable candidate habitable planet. Future observations as those discussed in Paper~II 
will allow us to further characterize Proxima b (e.g., spectral properties, orbital 
phase variations) and investigate the hypothetical presence of a thick atmosphere, which 
would open the exciting possibility for the planet hosting a liquid water reservoir. If 
this turned out to be the case, it would be noteworthy that the nearest star to the Sun 
also hosts the nearest habitable (perhaps inhabited?) planet.

\begin{acknowledgements}
B.~E. Wood is thanked for kindly providing the reduced HST/STIS H Ly $\alpha$ spectrum and model
for Proxima. I.~R. acknowledges support by the Spanish Ministry of Economy and Competitiveness 
(MINECO) through grant ESP2014-57495-C2-2-R. E.~B. acknowledges that this work is part 
of the F.R.S.-FNRS ``ExtraOrDynHa'' research project. 
The work of E. B. was supported by the Hubert Curien Tournesol Program.
E.~F.~G and S.~E. acknowledge support 
by the NSF and NASA through grants NSF/RUI-1009903, HST-GO-13020.001-A and Chandra Award 
GO2-13020X to Villanova University. S.~N.~R. acknowledges support from the Agence Nationale de 
la Recherche via grant ANR-13-BS05-0003-002 (project {\it MOJO}). We are grateful to the 
referees, Dr. Jason W. Barnes and an anonymous referee, for constructive, insightful, and 
timely reports.  
\end{acknowledgements}

   \bibliographystyle{aa} 
   \bibliography{biblio} 

\begin{thebibliography}{169}
\expandafter\ifx\csname natexlab\endcsname\relax\def\natexlab#1{#1}\fi

\bibitem[{{Airapetian} \& {Usmanov}(2016)}]{AirapetianUsmanov2016}
{Airapetian}, V.~S. \& {Usmanov}, A.~V. 2016, \apjl, 817, L24

\bibitem[{{Altwegg} {et~al.}(2015){Altwegg}, {Balsiger}, {Bar-Nun},
  {Berthelier}, {Bieler}, {Bochsler}, {Briois}, {Calmonte}, {Combi}, {De
  Keyser}, {Eberhardt}, {Fiethe}, {Fuselier}, {Gasc}, {Gombosi}, {Hansen},
  {H{\"a}ssig}, {J{\"a}ckel}, {Kopp}, {Korth}, {LeRoy}, {Mall}, {Marty},
  {Mousis}, {Neefs}, {Owen}, {R{\`e}me}, {Rubin}, {S{\'e}mon}, {Tzou}, {Waite},
  \& {Wurz}}]{altwegg15}
{Altwegg}, K., {Balsiger}, H., {Bar-Nun}, A., {et~al.} 2015, Science, 347,
  1261952

\bibitem[{{Anglada-Escud{\'e}} {et~al.}(2016){Anglada-Escud{\'e}}, {Amado},
  {Barnes}, {Berdi{\~n}as}, {Butler}, {Coleman}, {de La Cueva}, {Dreizler},
  {Endl}, {Giesers}, {Jeffers}, {Jenkins}, {Jones}, {Kiraga}, {K{\"u}rster},
  {L{\'o}pez-Gonz{\'a}lez}, {Marvin}, {Morales}, {Morin}, {Nelson}, {Ortiz},
  {Ofir}, {Paardekooper}, {Reiners}, {Rodr{\'{\i}}guez},
  {Rodr{\'{\i}}guez-L{\'o}pez}, {Sarmiento}, {Strachan}, {Tsapras}, {Tuomi}, \&
  {Zechmeister}}]{Anglada16}
{Anglada-Escud{\'e}}, G., {Amado}, P.~J., {Barnes}, J., {et~al.} 2016, \nat,
  536, 437

\bibitem[{{Atreya}(1999)}]{Atreya1999}
{Atreya}, S.~K. 1999, EOS Transactions, 80, 320

\bibitem[{{Audard} {et~al.}(2000){Audard}, {G{\"u}del}, {Drake}, \&
  {Kashyap}}]{Audardetal2000}
{Audard}, M., {G{\"u}del}, M., {Drake}, J.~J., \& {Kashyap}, V.~L. 2000, \apj,
  541, 396

\bibitem[{{Ayres}(2010)}]{Ayres2010}
{Ayres}, T.~R. 2010, \apjs, 187, 149

\bibitem[{{Badro} {et~al.}(2014){Badro}, {Cote}, \& {Brodholt}}]{badro14}
{Badro}, J., {Cote}, A., \& {Brodholt}, J. 2014, PNAS

\bibitem[{{Baraffe} {et~al.}(2015){Baraffe}, {Homeier}, {Allard}, \&
  {Chabrier}}]{Baraffe2015}
{Baraffe}, I., {Homeier}, D., {Allard}, F., \& {Chabrier}, G. 2015, \aap, 577,
  A42

\bibitem[{{Barnes} \& {Heller}(2013)}]{BarnesHeller2013}
{Barnes}, R. \& {Heller}, R. 2013, Astrobiology, 13, 279

\bibitem[{{Batalha} {et~al.}(2013){Batalha}, {Rowe}, {Bryson}, {Barclay},
  {Burke}, {Caldwell}, {Christiansen}, {Mullally}, {Thompson}, {Brown},
  {Dupree}, {Fabrycky}, {Ford}, {Fortney}, {Gilliland}, {Isaacson}, {Latham},
  {Marcy}, {Quinn}, {Ragozzine}, {Shporer}, {Borucki}, {Ciardi}, {Gautier},
  {Haas}, {Jenkins}, {Koch}, {Lissauer}, {Rapin}, {Basri}, {Boss}, {Buchhave},
  {Carter}, {Charbonneau}, {Christensen-Dalsgaard}, {Clarke}, {Cochran},
  {Demory}, {Desert}, {Devore}, {Doyle}, {Esquerdo}, {Everett}, {Fressin},
  {Geary}, {Girouard}, {Gould}, {Hall}, {Holman}, {Howard}, {Howell},
  {Ibrahim}, {Kinemuchi}, {Kjeldsen}, {Klaus}, {Li}, {Lucas}, {Meibom},
  {Morris}, {Pr{\v s}a}, {Quintana}, {Sanderfer}, {Sasselov}, {Seader},
  {Smith}, {Steffen}, {Still}, {Stumpe}, {Tarter}, {Tenenbaum}, {Torres},
  {Twicken}, {Uddin}, {Van Cleve}, {Walkowicz}, \& {Welsh}}]{batalha13}
{Batalha}, N.~M., {Rowe}, J.~F., {Bryson}, S.~T., {et~al.} 2013, \apjs, 204, 24

\bibitem[{{Bazot} {et~al.}(2016){Bazot}, {Christensen-Dalsgaard}, {Gizon}, \&
  {Benomar}}]{Bazotetal2016}
{Bazot}, M., {Christensen-Dalsgaard}, J., {Gizon}, L., \& {Benomar}, O. 2016,
  \mnras, 460, 1254

\bibitem[{{Birk} {et~al.}(2004){Birk}, {Lesch}, \& {Konz}}]{Birk2004}
{Birk}, G.~T., {Lesch}, H., \& {Konz}, C. 2004, \aap, 420, L15

\bibitem[{{Biver} {et~al.}(2016){Biver}, {Moreno}, {Bockel{\'e}e-Morvan},
  {Sandqvist}, {Colom}, {Crovisier}, {Lis}, {Boissier}, {Debout}, {Paubert},
  {Milam}, {Hjalmarson}, {Lundin}, {Karlsson}, {Battelino}, {Frisk}, {Murtagh},
  \& {Odin Team}}]{biver16}
{Biver}, N., {Moreno}, R., {Bockel{\'e}e-Morvan}, D., {et~al.} 2016, \aap, 589,
  A78

\bibitem[{{Bolmont} {et~al.}(2011){Bolmont}, {Raymond}, \&
  {Leconte}}]{Bolmont2011}
{Bolmont}, E., {Raymond}, S.~N., \& {Leconte}, J. 2011, A \& A, 535, A94

\bibitem[{{Bolmont} {et~al.}(2012){Bolmont}, {Raymond}, {Leconte}, \&
  {Matt}}]{Bolmont2012}
{Bolmont}, E., {Raymond}, S.~N., {Leconte}, J., \& {Matt}, S.~P. 2012, A\&A,
  544, A124

\bibitem[{{Bolmont} {et~al.}(2016){Bolmont}, {Selsis}, {Owen}, {Ribas},
  {Raymond}, {Leconte}, \& {Gillon}}]{Bolmont2016}
{Bolmont}, E., {Selsis}, F., {Owen}, J.~E., {et~al.} 2016, ArXiv e-prints

\bibitem[{{Borucki} {et~al.}(2010){Borucki}, {Koch}, {Basri}, {Batalha},
  {Brown}, {Caldwell}, {Caldwell}, {Christensen-Dalsgaard}, {Cochran},
  {DeVore}, {Dunham}, {Dupree}, {Gautier}, {Geary}, {Gilliland}, {Gould},
  {Howell}, {Jenkins}, {Kondo}, {Latham}, {Marcy}, {Meibom}, {Kjeldsen},
  {Lissauer}, {Monet}, {Morrison}, {Sasselov}, {Tarter}, {Boss}, {Brownlee},
  {Owen}, {Buzasi}, {Charbonneau}, {Doyle}, {Fortney}, {Ford}, {Holman},
  {Seager}, {Steffen}, {Welsh}, {Rowe}, {Anderson}, {Buchhave}, {Ciardi},
  {Walkowicz}, {Sherry}, {Horch}, {Isaacson}, {Everett}, {Fischer}, {Torres},
  {Johnson}, {Endl}, {MacQueen}, {Bryson}, {Dotson}, {Haas}, {Kolodziejczak},
  {Van Cleve}, {Chandrasekaran}, {Twicken}, {Quintana}, {Clarke}, {Allen},
  {Li}, {Wu}, {Tenenbaum}, {Verner}, {Bruhweiler}, {Barnes}, \&
  {Prsa}}]{borucki10}
{Borucki}, W.~J., {Koch}, D., {Basri}, G., {et~al.} 2010, Science, 327, 977

\bibitem[{{Boyajian} {et~al.}(2012){Boyajian}, {von Braun}, {van Belle},
  {McAlister}, {ten Brummelaar}, {Kane}, {Muirhead}, {Jones}, {White},
  {Schaefer}, {Ciardi}, {Henry}, {L{\'o}pez-Morales}, {Ridgway}, {Gies}, {Jao},
  {Rojas-Ayala}, {Parks}, {Sturmann}, {Sturmann}, {Turner}, {Farrington},
  {Goldfinger}, \& {Berger}}]{Boyajian2012}
{Boyajian}, T.~S., {von Braun}, K., {van Belle}, G., {et~al.} 2012, \apj, 757,
  112

\bibitem[{{Chamberlain} \& {Hunten}(1987)}]{Chamberlain_Hunten1987}
{Chamberlain}, J.~W. \& {Hunten}, D.~M. 1987, Orlando FL Academic Press Inc
  International Geophysics Series, 36

\bibitem[{{Christian} {et~al.}(2004){Christian}, {Mathioudakis}, {Bloomfield},
  {Dupuis}, \& {Keenan}}]{Christianetal2004}
{Christian}, D.~J., {Mathioudakis}, M., {Bloomfield}, D.~S., {Dupuis}, J., \&
  {Keenan}, F.~P. 2004, \apj, 612, 1140

\bibitem[{{Ciesla} {et~al.}(2015){Ciesla}, {Mulders}, {Pascucci}, \&
  {Apai}}]{ciesla15}
{Ciesla}, F.~J., {Mulders}, G.~D., {Pascucci}, I., \& {Apai}, D. 2015, \apj,
  804, 9

\bibitem[{{Claire} {et~al.}(2012){Claire}, {Sheets}, {Cohen}, {Ribas},
  {Meadows}, \& {Catling}}]{Claireetal2012}
{Claire}, M.~W., {Sheets}, J., {Cohen}, M., {et~al.} 2012, \apj, 757, 95

\bibitem[{{Correia} \& {Laskar}(2001)}]{CL01}
{Correia}, A.~C.~M. \& {Laskar}, J. 2001, \nat, 411, 767

\bibitem[{{Cossou} {et~al.}(2014){Cossou}, {Raymond}, {Hersant}, \&
  {Pierens}}]{cossou14}
{Cossou}, C., {Raymond}, S.~N., {Hersant}, F., \& {Pierens}, A. 2014, \aap,
  569, A56

\bibitem[{{Darwin}(1880)}]{Dar1880}
{Darwin}, G.~H. 1880, Royal Society of London Philosophical Transactions Series
  I, 171, 713

\bibitem[{{Delfosse} {et~al.}(2000){Delfosse}, {Forveille}, {S{\'e}gransan},
  {Beuzit}, {Udry}, {Perrier}, \& {Mayor}}]{Delfosse2000}
{Delfosse}, X., {Forveille}, T., {S{\'e}gransan}, D., {et~al.} 2000, \aap, 364,
  217

\bibitem[{{DeMeo} \& {Carry}(2013)}]{demeo13}
{DeMeo}, F. \& {Carry}, B. 2013, ArXiv e-prints

\bibitem[{{Demory} {et~al.}(2009){Demory}, {S{\'e}gransan}, {Forveille},
  {Queloz}, {Beuzit}, {Delfosse}, {di Folco}, {Kervella}, {Le Bouquin},
  {Perrier}, {Benisty}, {Duvert}, {Hofmann}, {Lopez}, \& {Petrov}}]{Demory2009}
{Demory}, B.-O., {S{\'e}gransan}, D., {Forveille}, T., {et~al.} 2009, \aap,
  505, 205

\bibitem[{{Desch} \& {Leshin}(2004)}]{desch04}
{Desch}, S.~J. \& {Leshin}, L.~A. 2004, in Lunar and Planetary Science
  Conference, Vol.~35, Lunar and Planetary Science Conference, ed.
  S.~{Mackwell} \& E.~{Stansbery}

\bibitem[{{do Nascimento} {et~al.}(2016){do Nascimento}, {Vidotto}, {Petit},
  {Folsom}, {Castro}, {Marsden}, {Morin}, {Porto de Mello}, {Meibom},
  {Jeffers}, {Guinan}, \& {Ribas}}]{doNascimentoetal2016}
{do Nascimento}, Jr., J.-D., {Vidotto}, A.~A., {Petit}, P., {et~al.} 2016,
  \apjl, 820, L15

\bibitem[{{Drake} {et~al.}(2015){Drake}, {Behar}, {Doyle}, {G{\"u}del},
  {Hamaguchi}, {Kowalski}, {Maccarone}, {Osten}, {Peretz}, \&
  {Wolk}}]{Drakeetal2015}
{Drake}, S.~A., {Behar}, E., {Doyle}, J.~G., {et~al.} 2015, ArXiv e-prints

\bibitem[{{Drake} {et~al.}(1996){Drake}, {Singh}, {White}, {Mewe}, \&
  {Kaastra}}]{Drakeetal1996}
{Drake}, S.~A., {Singh}, K.~P., {White}, N.~E., {Mewe}, R., \& {Kaastra}, J.~S.
  1996, in Astronomical Society of the Pacific Conference Series, Vol. 109,
  Cool Stars, Stellar Systems, and the Sun, ed. R.~{Pallavicini} \& A.~K.
  {Dupree}, 263

\bibitem[{{Efroimsky}(2012)}]{Efr12}
{Efroimsky}, M. 2012, \apj, 746, 150

\bibitem[{{Eggleton} {et~al.}(1998){Eggleton}, {Kiseleva}, \& {Hut}}]{EKH1998}
{Eggleton}, P.~P., {Kiseleva}, L.~G., \& {Hut}, P. 1998, ApJ, 499, 853

\bibitem[{{Elkins-Tanton}(2008)}]{2008E&PSL.271..181E}
{Elkins-Tanton}, L.~T. 2008, Earth and Planetary Science Letters, 271, 181

\bibitem[{{Engle} \& {Guinan}(2011)}]{EngleGuinan2011}
{Engle}, S.~G. \& {Guinan}, E.~F. 2011, in Astronomical Society of the Pacific
  Conference Series, Vol. 451, 9th Pacific Rim Conference on Stellar
  Astrophysics, ed. S.~{Qain}, K.~{Leung}, L.~{Zhu}, \& S.~{Kwok}, 285

\bibitem[{{Fabrycky} {et~al.}(2007){Fabrycky}, {Johnson}, \& {Goodman}}]{FJG07}
{Fabrycky}, D.~C., {Johnson}, E.~T., \& {Goodman}, J. 2007, \apj, 665, 754

\bibitem[{{Fortney} {et~al.}(2007){Fortney}, {Marley}, \&
  {Barnes}}]{Fortney2007}
{Fortney}, J.~J., {Marley}, M.~S., \& {Barnes}, J.~W. 2007, \apj, 659, 1661

\bibitem[{{Fuhrmeister} {et~al.}(2011){Fuhrmeister}, {Lalitha}, {Poppenhaeger},
  {Rudolf}, {Liefke}, {Reiners}, {Schmitt}, \& {Ness}}]{Fuhrmeisteretal2011}
{Fuhrmeister}, B., {Lalitha}, S., {Poppenhaeger}, K., {et~al.} 2011, \aap, 534,
  A133

\bibitem[{{Garaud} \& {Lin}(2007)}]{garaud07}
{Garaud}, P. \& {Lin}, D.~N.~C. 2007, \apj, 654, 606

\bibitem[{{Genda} \& {Abe}(2005)}]{genda05}
{Genda}, H. \& {Abe}, Y. 2005, \nat, 433, 842

\bibitem[{{Gerkema} {et~al.}(2004){Gerkema}, {Lam}, \&
  {Maas}}]{GerkemaLamMaas2004}
{Gerkema}, T., {Lam}, F.~A., \& {Maas}, L.~R.~M. 2004, Deep Sea Research Part
  II: Topical Studies in Oceanography, 51, 2995

\bibitem[{{Gillon} {et~al.}(2016){Gillon}, {Jehin}, {Lederer}, {Delrez}, {de
  Wit}, {Burdanov}, {Van Grootel}, {Burgasser}, {Triaud}, {Opitom}, {Demory},
  {Sahu}, {Bardalez Gagliuffi}, {Magain}, \& {Queloz}}]{Gillon2016}
{Gillon}, M., {Jehin}, E., {Lederer}, S.~M., {et~al.} 2016, \nat, 533, 221

\bibitem[{{Gold} \& {Soter}(1969)}]{GS69}
{Gold}, T. \& {Soter}, S. 1969, \icarus, 11, 356

\bibitem[{{Goldreich}(1963)}]{Gol63}
{Goldreich}, P. 1963, \mnras, 126, 257

\bibitem[{{Goldreich} \& {Peale}(1966)}]{GP66}
{Goldreich}, P. \& {Peale}, S. 1966, \aj, 71, 425

\bibitem[{{Goldreich} \& {Tremaine}(1980)}]{goldreich80}
{Goldreich}, P. \& {Tremaine}, S. 1980, \apj, 241, 425

\bibitem[{{Gomes} {et~al.}(2005){Gomes}, {Levison}, {Tsiganis}, \&
  {Morbidelli}}]{gomes05}
{Gomes}, R., {Levison}, H.~F., {Tsiganis}, K., \& {Morbidelli}, A. 2005, \nat,
  435, 466

\bibitem[{{Gradie} \& {Tedesco}(1982)}]{gradie82}
{Gradie}, J. \& {Tedesco}, E. 1982, Science, 216, 1405

\bibitem[{{Grimm} \& {McSween}(1993)}]{grimm93}
{Grimm}, R.~E. \& {McSween}, H.~Y. 1993, Science, 259, 653

\bibitem[{{G{\"u}del} {et~al.}(2004){G{\"u}del}, {Audard}, {Reale}, {Skinner},
  \& {Linsky}}]{Guedeletal2004}
{G{\"u}del}, M., {Audard}, M., {Reale}, F., {Skinner}, S.~L., \& {Linsky},
  J.~L. 2004, \aap, 416, 713

\bibitem[{{Guinan} {et~al.}(2016){Guinan}, {Engle}, \&
  {Durbin}}]{Guinanetal2016}
{Guinan}, E.~F., {Engle}, S.~G., \& {Durbin}, A. 2016, \apj, 821, 81

\bibitem[{{Guinan} {et~al.}(2003){Guinan}, {Ribas}, \&
  {Harper}}]{Guinanetal2003}
{Guinan}, E.~F., {Ribas}, I., \& {Harper}, G.~M. 2003, \apj, 594, 561

\bibitem[{{Haisch} {et~al.}(1983){Haisch}, {Linsky}, {Bornmann}, {Stencel},
  {Antiochos}, {Golub}, \& {Vaiana}}]{Haischetal1983}
{Haisch}, B.~M., {Linsky}, J.~L., {Bornmann}, P.~L., {et~al.} 1983, \apj, 267,
  280

\bibitem[{{Haisch} {et~al.}(2001){Haisch}, {Lada}, \& {Lada}}]{haisch01}
{Haisch}, Jr., K.~E., {Lada}, E.~A., \& {Lada}, C.~J. 2001, \apjl, 553, L153

\bibitem[{{Hamano} {et~al.}(2013){Hamano}, {Abe}, \& {Genda}}]{Hamano2013}
{Hamano}, K., {Abe}, Y., \& {Genda}, H. 2013, \nat, 497, 607

\bibitem[{{Hansen}(2010)}]{Hansen2010}
{Hansen}, B.~M.~S. 2010, ApJ, 723, 285

\bibitem[{{Hansen}(2015)}]{hansen15}
{Hansen}, B.~M.~S. 2015, International Journal of Astrobiology, 14, 267

\bibitem[{{Hartogh} {et~al.}(2011){Hartogh}, {Lis}, {Bockel{\'e}e-Morvan}, {de
  Val-Borro}, {Biver}, {K{\"u}ppers}, {Emprechtinger}, {Bergin}, {Crovisier},
  {Rengel}, {Moreno}, {Szutowicz}, \& {Blake}}]{hartogh11}
{Hartogh}, P., {Lis}, D.~C., {Bockel{\'e}e-Morvan}, D., {et~al.} 2011, \nat,
  478, 218

\bibitem[{{Hudson}(1991)}]{Hudson1991}
{Hudson}, H.~S. 1991, \solphys, 133, 357

\bibitem[{{H{\"u}nsch} {et~al.}(1999){H{\"u}nsch}, {Schmitt}, {Sterzik}, \&
  {Voges}}]{Hunschetal1999}
{H{\"u}nsch}, M., {Schmitt}, J.~H.~M.~M., {Sterzik}, M.~F., \& {Voges}, W.
  1999, \aaps, 135, 319

\bibitem[{{Hunten} {et~al.}(1987){Hunten}, {Pepin}, \& {Walker}}]{Hunten1987}
{Hunten}, D.~M., {Pepin}, R.~O., \& {Walker}, J.~C.~G. 1987, \icarus, 69, 532

\bibitem[{{Hut}(1980)}]{Hut80}
{Hut}, P. 1980, \aap, 92, 167

\bibitem[{{Hut}(1981)}]{Hut1981}
{Hut}, P. 1981, A \& A, 99, 126

\bibitem[{{Ida} \& {Lin}(2010)}]{ida10}
{Ida}, S. \& {Lin}, D.~N.~C. 2010, \apj, 719, 810

\bibitem[{{Ingersoll} \& {Dobrovolskis}(1978)}]{ID78}
{Ingersoll}, A.~P. \& {Dobrovolskis}, A.~R. 1978, \nat, 275, 37

\bibitem[{{Izidoro} \& {et al.}(2016)}]{izidoro16}
{Izidoro}, A. \& {et al.} 2016, Submitted to MNRAS

\bibitem[{{Izidoro} {et~al.}(2015){Izidoro}, {Raymond}, {Morbidelli}, \&
  {Winter}}]{izidoro15}
{Izidoro}, A., {Raymond}, S.~N., {Morbidelli}, A., \& {Winter}, O.~C. 2015,
  \mnras, 453, 3619

\bibitem[{{Jardine} \& {Unruh}(1999)}]{JardineUnruh1999}
{Jardine}, M. \& {Unruh}, Y.~C. 1999, \aap, 346, 883

\bibitem[{{Kaib} {et~al.}(2013){Kaib}, {Raymond}, \& {Duncan}}]{Kaib2013}
{Kaib}, N.~A., {Raymond}, S.~N., \& {Duncan}, M. 2013, \nat, 493, 381

\bibitem[{{Kalyaan} \& {Desch}(2016)}]{kalyaan16}
{Kalyaan}, A. \& {Desch}, S. 2016, in American Astronomical Society Meeting
  Abstracts, Vol. 228, American Astronomical Society Meeting Abstracts, 320.05

\bibitem[{{Kasting} {et~al.}(1993){Kasting}, {Whitmire}, \&
  {Reynolds}}]{Kasting1993}
{Kasting}, J.~F., {Whitmire}, D.~P., \& {Reynolds}, R.~T. 1993, Icarus, 101,
  108

\bibitem[{{Kennedy} \& {Kenyon}(2008)}]{kennedy08}
{Kennedy}, G.~M. \& {Kenyon}, S.~J. 2008, \apj, 673, 502

\bibitem[{{Khodachenko} {et~al.}(2012){Khodachenko}, {Alexeev}, {Belenkaya},
  {Lammer}, {Grie{\ss}meier}, {Leitzinger}, {Odert}, {Zaqarashvili}, \&
  {Rucker}}]{Khodachenko2012}
{Khodachenko}, M.~L., {Alexeev}, I., {Belenkaya}, E., {et~al.} 2012, \apj, 744,
  70

\bibitem[{{Khodachenko} {et~al.}(2007){Khodachenko}, {Ribas}, {Lammer},
  {Grie{\ss}meier}, {Leitner}, {Selsis}, {Eiroa}, {Hanslmeier}, {Biernat},
  {Farrugia}, \& {Rucker}}]{Khodachenko2007}
{Khodachenko}, M.~L., {Ribas}, I., {Lammer}, H., {et~al.} 2007, Astrobiology,
  7, 167

\bibitem[{{Kopparapu}(2013)}]{Kopparapu2013}
{Kopparapu}, R.~K. 2013, ApJL, 767, L8

\bibitem[{{Kopparapu} {et~al.}(2014){Kopparapu}, {Ramirez}, {SchottelKotte},
  {Kasting}, {Domagal-Goldman}, \& {Eymet}}]{Kopparapu2014}
{Kopparapu}, R.~K., {Ramirez}, R.~M., {SchottelKotte}, J., {et~al.} 2014,
  \apjl, 787, L29

\bibitem[{{Kopparapu} {et~al.}(2016){Kopparapu}, {Wolf}, {Haqq-Misra}, {Yang},
  {Kasting}, {Meadows}, {Terrien}, \& {Mahadevan}}]{Kopparapu2016}
{Kopparapu}, R.~k., {Wolf}, E.~T., {Haqq-Misra}, J., {et~al.} 2016, \apj, 819,
  84

\bibitem[{{Kuchner}(2003)}]{kuchner03}
{Kuchner}, M.~J. 2003, \apjl, 596, L105

\bibitem[{{Kunkel}(1973)}]{Kunkel1973}
{Kunkel}, W.~E. 1973, \apjs, 25, 1

\bibitem[{{Lammer} {et~al.}(2009){Lammer}, {Bredeh{\"o}ft}, {Coustenis},
  {Khodachenko}, {Kaltenegger}, {Grasset}, {Prieur}, {Raulin}, {Ehrenfreund},
  {Yamauchi}, {Wahlund}, {Grie{\ss}meier}, {Stangl}, {Cockell}, {Kulikov},
  {Grenfell}, \& {Rauer}}]{Lammeretal2009}
{Lammer}, H., {Bredeh{\"o}ft}, J.~H., {Coustenis}, A., {et~al.} 2009, \aapr,
  17, 181

\bibitem[{{Lammer} {et~al.}(2012){Lammer}, {G{\"u}del}, {Kulikov}, {Ribas},
  {Zaqarashvili}, {Khodachenko}, {Kislyakova}, {Gr{\"o}ller}, {Odert},
  {Leitzinger}, {Fichtinger}, {Krauss}, {Hausleitner}, {Holmstr{\"o}m},
  {Sanz-Forcada}, {Lichtenegger}, {Hanslmeier}, {Shematovich}, {Bisikalo},
  {Rauer}, \& {Fridlund}}]{Lammer2012}
{Lammer}, H., {G{\"u}del}, M., {Kulikov}, Y., {et~al.} 2012, Earth, Planets,
  and Space, 64, 179

\bibitem[{{Lammer} {et~al.}(2011){Lammer}, {Kislyakova}, {Odert}, {Leitzinger},
  {Schwarz}, {Pilat-Lohinger}, {Kulikov}, {Khodachenko}, {G{\"u}del}, \&
  {Hanslmeier}}]{Lammer2011}
{Lammer}, H., {Kislyakova}, K.~G., {Odert}, P., {et~al.} 2011, Origins of Life
  and Evolution of the Biosphere, 41, 503

\bibitem[{{Lammer} {et~al.}(2007){Lammer}, {Lichtenegger}, {Kulikov},
  {Grie{\ss}meier}, {Terada}, {Erkaev}, {Biernat}, {Khodachenko}, {Ribas},
  {Penz}, \& {Selsis}}]{Lammer2007}
{Lammer}, H., {Lichtenegger}, H.~I.~M., {Kulikov}, Y.~N., {et~al.} 2007,
  Astrobiology, 7, 185

\bibitem[{{Lammer} {et~al.}(2003){Lammer}, {Selsis}, {Ribas}, {Guinan},
  {Bauer}, \& {Weiss}}]{Lammer2003}
{Lammer}, H., {Selsis}, F., {Ribas}, I., {et~al.} 2003, ApJl, 598, L121

\bibitem[{{Lecar} {et~al.}(2006){Lecar}, {Podolak}, {Sasselov}, \&
  {Chiang}}]{lecar06}
{Lecar}, M., {Podolak}, M., {Sasselov}, D., \& {Chiang}, E. 2006, \apj, 640,
  1115

\bibitem[{{Leconte} {et~al.}(2015){Leconte}, {Wu}, {Menou}, \&
  {Murray}}]{LWM15}
{Leconte}, J., {Wu}, H., {Menou}, K., \& {Murray}, N. 2015, Science, 347, 632

\bibitem[{{L{\'e}cuyer} {et~al.}(1998){L{\'e}cuyer}, {Gillet}, \&
  {Robert}}]{lecuyer98}
{L{\'e}cuyer}, C., {Gillet}, P., \& {Robert}, F. 1998, Chem. Geol., 145, 249

\bibitem[{{L{\'e}ger} {et~al.}(2004){L{\'e}ger}, {Selsis}, {Sotin}, {Guillot},
  {Despois}, {Mawet}, {Ollivier}, {Lab{\`e}que}, {Valette}, {Brachet},
  {Chazelas}, \& {Lammer}}]{leger04}
{L{\'e}ger}, A., {Selsis}, F., {Sotin}, C., {et~al.} 2004, \icarus, 169, 499

\bibitem[{{Lichtenegger} {et~al.}(2010){Lichtenegger}, {Lammer},
  {Grie{\ss}meier}, {Kulikov}, {von Paris}, {Hausleitner}, {Krauss}, \&
  {Rauer}}]{Lichtenegger2010}
{Lichtenegger}, H.~I.~M., {Lammer}, H., {Grie{\ss}meier}, J.-M., {et~al.} 2010,
  \icarus, 210, 1

\bibitem[{{Linsky} {et~al.}(2014){Linsky}, {Fontenla}, \&
  {France}}]{Linskyetal2014}
{Linsky}, J.~L., {Fontenla}, J., \& {France}, K. 2014, \apj, 780, 61

\bibitem[{{Linsky} \& {Wood}(2014)}]{LinskyWood2014}
{Linsky}, J.~L. \& {Wood}, B.~E. 2014, ASTRA Proceedings, 1, 43

\bibitem[{{Lis} {et~al.}(2013){Lis}, {Biver}, {Bockel{\'e}e-Morvan}, {Hartogh},
  {Bergin}, {Blake}, {Crovisier}, {de Val-Borro}, {Jehin}, {K{\"u}ppers},
  {Manfroid}, {Moreno}, {Rengel}, \& {Szutowicz}}]{lis13}
{Lis}, D.~C., {Biver}, N., {Bockel{\'e}e-Morvan}, D., {et~al.} 2013, \apjl,
  774, L3

\bibitem[{{Lissauer}(2007)}]{lissauer07}
{Lissauer}, J.~J. 2007, \apjl, 660, L149

\bibitem[{{Love}(1909)}]{Lov09}
{Love}, A.~E.~H. 1909, \mnras, 69, 476

\bibitem[{{Loyd} \& {France}(2014)}]{LoydFrance2014}
{Loyd}, R.~O.~P. \& {France}, K. 2014, \apjs, 211, 9

\bibitem[{{Luger} \& {Barnes}(2015)}]{LugerBarnes2015}
{Luger}, R. \& {Barnes}, R. 2015, Astrobiology, 15, 119

\bibitem[{{Makarov}(2012)}]{Mak12}
{Makarov}, V.~V. 2012, \apj, 752, 73

\bibitem[{{Makarov} \& {Efroimsky}(2013)}]{ME13}
{Makarov}, V.~V. \& {Efroimsky}, M. 2013, \apj, 764, 27

\bibitem[{{Martin} \& {Livio}(2012)}]{martin12}
{Martin}, R.~G. \& {Livio}, M. 2012, \mnras, 425, L6

\bibitem[{{Marty}(2012{\natexlab{a}})}]{marty12}
{Marty}, B. 2012{\natexlab{a}}, Earth and Planetary Science Letters, 313, 56

\bibitem[{{Marty}(2012{\natexlab{b}})}]{Marty2012}
{Marty}, B. 2012{\natexlab{b}}, Earth and Planetary Science Letters, 313, 56

\bibitem[{{Marty} {et~al.}(2016){Marty}, {Avice}, {Sano}, {Altwegg},
  {Balsiger}, {H{\"a}ssig}, {Morbidelli}, {Mousis}, \& {Rubin}}]{marty16}
{Marty}, B., {Avice}, G., {Sano}, Y., {et~al.} 2016, Earth and Planetary
  Science Letters, 441, 91

\bibitem[{{Marty} \& {Yokochi}(2006)}]{marty06}
{Marty}, B. \& {Yokochi}, R. 2006, Rev. Mineral Geophys., 62, 421

\bibitem[{{Mayor} {et~al.}(2011){Mayor}, {Marmier}, {Lovis}, {Udry},
  {S{\'e}gransan}, {Pepe}, {Benz}, {Bertaux}, {Bouchy}, {Dumusque}, {Lo Curto},
  {Mordasini}, {Queloz}, \& {Santos}}]{mayor11}
{Mayor}, M., {Marmier}, M., {Lovis}, C., {et~al.} 2011, ArXiv e-prints

\bibitem[{{McNeil} \& {Nelson}(2010)}]{mcneil10}
{McNeil}, D.~S. \& {Nelson}, R.~P. 2010, \mnras, 401, 1691

\bibitem[{{Mignard}(1979)}]{Mignard1979}
{Mignard}, F. 1979, Moon and Planets, 20, 301

\bibitem[{{Montgomery} \& {Laughlin}(2009)}]{montgomery09}
{Montgomery}, R. \& {Laughlin}, G. 2009, \icarus, 202, 1

\bibitem[{{Morbidelli} {et~al.}(2000){Morbidelli}, {Chambers}, {Lunine},
  {Petit}, {Robert}, {Valsecchi}, \& {Cyr}}]{morby00}
{Morbidelli}, A., {Chambers}, J., {Lunine}, J.~I., {et~al.} 2000, Meteoritics
  and Planetary Science, 35, 1309

\bibitem[{{Morbidelli} {et~al.}(2015){Morbidelli}, {Lambrechts}, {Jacobson}, \&
  {Bitsch}}]{morby16}
{Morbidelli}, A., {Lambrechts}, M., {Jacobson}, S., \& {Bitsch}, B. 2015,
  Icarus, 258, 418

\bibitem[{{Morbidelli} {et~al.}(2012){Morbidelli}, {Lunine}, {O'Brien},
  {Raymond}, \& {Walsh}}]{morby12}
{Morbidelli}, A., {Lunine}, J.~I., {O'Brien}, D.~P., {Raymond}, S.~N., \&
  {Walsh}, K.~J. 2012, Annual Review of Earth and Planetary Sciences, 40, 251

\bibitem[{{Morin} {et~al.}(2010){Morin}, {Donati}, {Petit}, {Delfosse},
  {Forveille}, \& {Jardine}}]{Morin10}
{Morin}, J., {Donati}, J.-F., {Petit}, P., {et~al.} 2010, \mnras, 407, 2269

\bibitem[{{Mulders} {et~al.}(2015){Mulders}, {Ciesla}, {Min}, \&
  {Pascucci}}]{mulders15}
{Mulders}, G.~D., {Ciesla}, F.~J., {Min}, M., \& {Pascucci}, I. 2015, \apj,
  807, 9

\bibitem[{{Neron de Surgy} \& {Laskar}(1997)}]{deSurgyLaskar1997}
{Neron de Surgy}, O. \& {Laskar}, J. 1997, A \& A, 318, 975

\bibitem[{{Neves} {et~al.}(2013){Neves}, {Bonfils}, {Santos}, {Delfosse},
  {Forveille}, {Allard}, \& {Udry}}]{Nevesetal2013}
{Neves}, V., {Bonfils}, X., {Santos}, N.~C., {et~al.} 2013, \aap, 551, A36

\bibitem[{{O'Brien} {et~al.}(2014){O'Brien}, {Walsh}, {Morbidelli}, {Raymond},
  \& {Mandell}}]{obrien14}
{O'Brien}, D.~P., {Walsh}, K.~J., {Morbidelli}, A., {Raymond}, S.~N., \&
  {Mandell}, A.~M. 2014, \icarus, 239, 74

\bibitem[{{Ogihara} \& {Ida}(2009)}]{ogihara09}
{Ogihara}, M. \& {Ida}, S. 2009, \apj, 699, 824

\bibitem[{{Owen} \& {Alvarez}(2016)}]{OwenAlvarez2016}
{Owen}, J.~E. \& {Alvarez}, M.~A. 2016, \apj, 816, 34

\bibitem[{{Paardekooper} {et~al.}(2011){Paardekooper}, {Baruteau}, \&
  {Kley}}]{paardekooper11}
{Paardekooper}, S.-J., {Baruteau}, C., \& {Kley}, W. 2011, \mnras, 410, 293

\bibitem[{{Panero}(2016)}]{panero16}
{Panero}, W. e.~a. 2016, Submitted to Nature Geosciences

\bibitem[{{Pascucci} {et~al.}(2009{\natexlab{a}}){Pascucci}, {Apai}, {Luhman},
  {Henning}, {Bouwman}, {Meyer}, {Lahuis}, \& {Natta}}]{pascucci09}
{Pascucci}, I., {Apai}, D., {Luhman}, K., {et~al.} 2009{\natexlab{a}}, \apj,
  696, 143

\bibitem[{{Pascucci} {et~al.}(2009{\natexlab{b}}){Pascucci}, {Apai}, {Luhman},
  {Henning}, {Bouwman}, {Meyer}, {Lahuis}, \& {Natta}}]{Pascucci2009}
{Pascucci}, I., {Apai}, D., {Luhman}, K., {et~al.} 2009{\natexlab{b}}, ApJ,
  696, 143

\bibitem[{{Pecaut} \& {Mamajek}(2016)}]{PecautMamajek2016}
{Pecaut}, M.~J. \& {Mamajek}, E.~E. 2016, ArXiv e-prints

\bibitem[{{Pfalzner} {et~al.}(2014){Pfalzner}, {Steinhausen}, \&
  {Menten}}]{Pfalzner2014}
{Pfalzner}, S., {Steinhausen}, M., \& {Menten}, K. 2014, \apjl, 793, L34

\bibitem[{{Podolak}(2010)}]{podolak10}
{Podolak}, M. 2010, in IAU Symposium, Vol. 263, IAU Symposium, ed. J.~A.
  {Fernandez}, D.~{Lazzaro}, D.~{Prialnik}, \& R.~{Schulz}, 19--28

\bibitem[{{Pollack} {et~al.}(1993){Pollack}, {Hurter}, \&
  {Johnson}}]{Pollack1993}
{Pollack}, H.~N., {Hurter}, S.~J., \& {Johnson}, J.~R. 1993, Reviews of
  Geophysics, 31, 267

\bibitem[{{Ramirez} \& {Kaltenegger}(2014)}]{RamirezKaltenegger2014}
{Ramirez}, R.~M. \& {Kaltenegger}, L. 2014, \apjl, 797, L25

\bibitem[{{Raymond} {et~al.}(2014){Raymond}, {Kokubo}, {Morbidelli},
  {Morishima}, \& {Walsh}}]{raymond14}
{Raymond}, S.~N., {Kokubo}, E., {Morbidelli}, A., {Morishima}, R., \& {Walsh},
  K.~J. 2014, Protostars and Planets VI, 595

\bibitem[{{Raymond} {et~al.}(2009){Raymond}, {O'Brien}, {Morbidelli}, \&
  {Kaib}}]{raymond09}
{Raymond}, S.~N., {O'Brien}, D.~P., {Morbidelli}, A., \& {Kaib}, N.~A. 2009,
  Icarus, 203, 644

\bibitem[{{Raymond} {et~al.}(2004){Raymond}, {Quinn}, \& {Lunine}}]{raymond04}
{Raymond}, S.~N., {Quinn}, T., \& {Lunine}, J.~I. 2004, Icarus, 168, 1

\bibitem[{{Raymond} {et~al.}(2006){Raymond}, {Quinn}, \& {Lunine}}]{raymond06}
{Raymond}, S.~N., {Quinn}, T., \& {Lunine}, J.~I. 2006, Icarus, 183, 265

\bibitem[{{Raymond} {et~al.}(2007{\natexlab{a}}){Raymond}, {Quinn}, \&
  {Lunine}}]{raymond07a}
{Raymond}, S.~N., {Quinn}, T., \& {Lunine}, J.~I. 2007{\natexlab{a}},
  Astrobiology, 7, 66

\bibitem[{{Raymond} {et~al.}(2007{\natexlab{b}}){Raymond}, {Scalo}, \&
  {Meadows}}]{raymond07b}
{Raymond}, S.~N., {Scalo}, J., \& {Meadows}, V.~S. 2007{\natexlab{b}}, \apj,
  669, 606

\bibitem[{{Reiners} \& {Basri}(2009)}]{Reiners09}
{Reiners}, A. \& {Basri}, G. 2009, \aap, 496, 787

\bibitem[{{Reiners} \& {Mohanty}(2012)}]{Reiners12}
{Reiners}, A. \& {Mohanty}, S. 2012, \apj, 746, 43

\bibitem[{{Reiners} {et~al.}(2014){Reiners}, {Sch{\"u}ssler}, \&
  {Passegger}}]{Reinersetal2014}
{Reiners}, A., {Sch{\"u}ssler}, M., \& {Passegger}, V.~M. 2014, \apj, 794, 144

\bibitem[{{Ribas} {et~al.}(2005){Ribas}, {Guinan}, {G{\"u}del}, \&
  {Audard}}]{Ribasetal2005}
{Ribas}, I., {Guinan}, E.~F., {G{\"u}del}, M., \& {Audard}, M. 2005, \apj, 622,
  680

\bibitem[{{Rosing}(2005)}]{Rosing2005}
{Rosing}, M.~T. 2005, International Journal of Astrobiology, 4, 9

\bibitem[{{Roskosz} {et~al.}(2013){Roskosz}, {Bouhifd}, {Jephcoat}, {Marty}, \&
  {Mysen}}]{Roskosz2013}
{Roskosz}, M., {Bouhifd}, M.~A., {Jephcoat}, A.~P., {Marty}, B., \& {Mysen},
  B.~O. 2013, \gca, 121, 15

\bibitem[{{Sasselov} \& {Lecar}(2000)}]{sasselov00}
{Sasselov}, D.~D. \& {Lecar}, M. 2000, \apj, 528, 995

\bibitem[{{Schmitt} \& {Liefke}(2004)}]{SchmittLiefke2004}
{Schmitt}, J.~H.~M.~M. \& {Liefke}, C. 2004, \aap, 417, 651

\bibitem[{{Selsis} {et~al.}(2007{\natexlab{a}}){Selsis}, {Chazelas},
  {Bord{\'e}}, {Ollivier}, {Brachet}, {Decaudin}, {Bouchy}, {Ehrenreich},
  {Griessmeier}, {Lammer}, {Sotin}, {Grasset}, {Moutou}, {Barge}, {Deleuil},
  {Mawet}, {Despois}, {Kasting}, \& {L{\'e}ger}}]{Selsis2007a}
{Selsis}, F., {Chazelas}, B., {Bord{\'e}}, P., {et~al.} 2007{\natexlab{a}},
  Icarus, 191, 453

\bibitem[{{Selsis} {et~al.}(2007{\natexlab{b}}){Selsis}, {Kasting}, {Levrard},
  {Paillet}, {Ribas}, \& {Delfosse}}]{Selsis2007}
{Selsis}, F., {Kasting}, J.~F., {Levrard}, B., {et~al.} 2007{\natexlab{b}}, A
  \& A, 476, 1373

\bibitem[{{Semel}(1989)}]{Semel89}
{Semel}, M. 1989, \aap, 225, 456

\bibitem[{{Shibayama} {et~al.}(2013){Shibayama}, {Maehara}, {Notsu}, {Notsu},
  {Nagao}, {Honda}, {Ishii}, {Nogami}, \& {Shibata}}]{Shibayamaetal2013}
{Shibayama}, T., {Maehara}, H., {Notsu}, S., {et~al.} 2013, \apjs, 209, 5

\bibitem[{{Smith} {et~al.}(2012){Smith}, {Zuber}, {Phillips}, {Solomon},
  {Hauck}, {Lemoine}, {Mazarico}, {Neumann}, {Peale}, {Margot}, {Johnson},
  {Torrence}, {Perry}, {Rowlands}, {Goossens}, {Head}, \& {Taylor}}]{SZP12}
{Smith}, D.~E., {Zuber}, M.~T., {Phillips}, R.~J., {et~al.} 2012, Science, 336,
  214

\bibitem[{{Sok{\'o}{\l}} {et~al.}(2013){Sok{\'o}{\l}}, {Bzowski}, {Tokumaru},
  {Fujiki}, \& {McComas}}]{Sokoletal2013}
{Sok{\'o}{\l}}, J.~M., {Bzowski}, M., {Tokumaru}, M., {Fujiki}, K., \&
  {McComas}, D.~J. 2013, \solphys, 285, 167

\bibitem[{{Spencer} {et~al.}(2000){Spencer}, {Rathbun}, {Travis}, {Tamppari},
  {Barnard}, {Martin}, \& {McEwen}}]{Spencer2000}
{Spencer}, J.~R., {Rathbun}, J.~A., {Travis}, L.~D., {et~al.} 2000, Science,
  288, 1198

\bibitem[{{Swift} {et~al.}(2013){Swift}, {Johnson}, {Morton}, {Crepp},
  {Montet}, {Fabrycky}, \& {Muirhead}}]{swift13}
{Swift}, J.~J., {Johnson}, J.~A., {Morton}, T.~D., {et~al.} 2013, \apj, 764,
  105

\bibitem[{{Terquem} \& {Papaloizou}(2007)}]{terquem07}
{Terquem}, C. \& {Papaloizou}, J.~C.~B. 2007, \apj, 654, 1110

\bibitem[{{Thuillier} {et~al.}(2004){Thuillier}, {Floyd}, {Woods}, {Cebula},
  {Hilsenrath}, {Hers{\'e}}, \& {Labs}}]{Thuillieretal2004}
{Thuillier}, G., {Floyd}, L., {Woods}, T.~N., {et~al.} 2004, in Washington DC
  American Geophysical Union Geophysical Monograph Series, Vol. 141, Solar
  Variability and its Effects on Climate. Geophysical Monograph 141, ed. J.~M.
  {Pap}, P.~{Fox}, C.~{Frohlich}, H.~S. {Hudson}, J.~{Kuhn}, J.~{McCormack},
  G.~{North}, W.~{Sprigg}, \& S.~T. {Wu}, 171

\bibitem[{{Tian}(2009)}]{Tian2009}
{Tian}, F. 2009, \apj, 703, 905

\bibitem[{{Tian} {et~al.}(2008){Tian}, {Kasting}, {Liu}, \&
  {Roble}}]{Tian2008_1}
{Tian}, F., {Kasting}, J.~F., {Liu}, H.-L., \& {Roble}, R.~G. 2008, Journal of
  Geophysical Research (Planets), 113, E05008

\bibitem[{{Tolstikhin} {et~al.}(2014){Tolstikhin}, {Marty}, {Porcelli}, \&
  {Hofmann}}]{Tolstikhin2014}
{Tolstikhin}, I., {Marty}, B., {Porcelli}, D., \& {Hofmann}, A. 2014, \gca,
  136, 229

\bibitem[{{Turbet} {et~al.}(2016){Turbet}, {Leconte}, {Selsis}, {Bolmont},
  {Forget}, \& {Ribas}}]{Turbet16}
{Turbet}, M., {Leconte}, J., {Selsis}, F., {et~al.} 2016, Astronomy and
  Astrophysics, submitted for publication

\bibitem[{{Van Laerhoven} {et~al.}(2014){Van Laerhoven}, {Barnes}, \&
  {Greenberg}}]{vanLaerhovenetal2014}
{Van Laerhoven}, C., {Barnes}, R., \& {Greenberg}, R. 2014, \mnras, 441, 1888

\bibitem[{{Vidotto} {et~al.}(2013){Vidotto}, {Jardine}, {Morin}, {Donati},
  {Lang}, \& {Russell}}]{Vidotto13}
{Vidotto}, A.~A., {Jardine}, M., {Morin}, J., {et~al.} 2013, \aap, 557, A67

\bibitem[{{Walker}(1981)}]{Walker1981}
{Walker}, A.~R. 1981, \mnras, 195, 1029

\bibitem[{{Walsh} {et~al.}(2011){Walsh}, {Morbidelli}, {Raymond}, {O'Brien}, \&
  {Mandell}}]{walsh11}
{Walsh}, K.~J., {Morbidelli}, A., {Raymond}, S.~N., {O'Brien}, D.~P., \&
  {Mandell}, A.~M. 2011, \nat, 475, 206

\bibitem[{{Wood} {et~al.}(2001){Wood}, {Linsky}, {M{\"u}ller}, \&
  {Zank}}]{Woodetal2001}
{Wood}, B.~E., {Linsky}, J.~L., {M{\"u}ller}, H.-R., \& {Zank}, G.~P. 2001,
  \apjl, 547, L49

\bibitem[{{Wood} {et~al.}(2014){Wood}, {M{\"u}ller}, {Redfield}, \&
  {Edelman}}]{Woodetal2014}
{Wood}, B.~E., {M{\"u}ller}, H.-R., {Redfield}, S., \& {Edelman}, E. 2014,
  \apjl, 781, L33

\bibitem[{{Wood} {et~al.}(2005){Wood}, {Redfield}, {Linsky}, {M{\"u}ller}, \&
  {Zank}}]{Woodetal2005}
{Wood}, B.~E., {Redfield}, S., {Linsky}, J.~L., {M{\"u}ller}, H.-R., \& {Zank},
  G.~P. 2005, \apjs, 159, 118

\bibitem[{{Worth} \& {Sigurdsson}(2016)}]{WorthSigurdsson2016}
{Worth}, R. \& {Sigurdsson}, S. 2016, ArXiv e-prints

\bibitem[{{Yang} {et~al.}(2013){Yang}, {Cowan}, \& {Abbot}}]{Yang2013}
{Yang}, J., {Cowan}, N.~B., \& {Abbot}, D.~S. 2013, \apjl, 771, L45

\bibitem[{{Yoder}(1995)}]{Yod95}
{Yoder}, C.~F. 1995, in Global Earth Physics: A Handbook of Physical Constants,
  ed. T.~J. {Ahrens}, 1

\bibitem[{{Zahnle}(2015)}]{Zahnle_LPI_2015}
{Zahnle}, K.~J. 2015, in Lunar and Planetary Inst.~Technical Report, Vol.~46,
  Lunar and Planetary Science Conference, 1549

\bibitem[{{Zechmeister} {et~al.}(2009){Zechmeister}, {K{\"u}rster}, \&
  {Endl}}]{Zechmeisteretal2009}
{Zechmeister}, M., {K{\"u}rster}, M., \& {Endl}, M. 2009, \aap, 505, 859

\bibitem[{{Zhang} {et~al.}(2007){Zhang}, {Delva}, {Baumjohann}, {Auster},
  {Carr}, {Russell}, {Barabash}, {Balikhin}, {Kudela}, {Berghofer}, {Biernat},
  {Lammer}, {Lichtenegger}, {Magnes}, {Nakamura}, {Schwingenschuh}, {Volwerk},
  {V{\"o}r{\"o}s}, {Zambelli}, {Fornacon}, {Glassmeier}, {Richter}, {Balogh},
  {Schwarzl}, {Pope}, {Shi}, {Wang}, {Motschmann}, \& {Lebreton}}]{Zhang2007}
{Zhang}, T.~L., {Delva}, M., {Baumjohann}, W., {et~al.} 2007, \nat, 450, 654

\bibitem[{{Zuluaga} {et~al.}(2013){Zuluaga}, {Bustamante}, {Cuartas}, \&
  {Hoyos}}]{Zuluaga13}
{Zuluaga}, J.~I., {Bustamante}, S., {Cuartas}, P.~A., \& {Hoyos}, J.~H. 2013,
  \apj, 770, 23

\end{thebibliography}

\appendix
\section{Magnetopause radius of Proxima~b} \label{app1}

\subsection{Case without wind ram pressure}

This is the approach of \citet[][hereafter V13]{Vidotto13}, which relies on the extrapolation 
of the large-scale surface magnetic field of a star -- derived by means of Zeeman-Doppler 
Imaging (ZDI) -- using the potential field source surface methods. The equation 2 of V13 
defining the pressure balance at the nose of the magnetopause then becomes:
\begin{equation}
  \dfrac{B_{ss}^2}{8\pi} \left(\dfrac{R_{ss}}{R_{orb}}\right)^4
  = \dfrac{\left(\dfrac{1}{2}B_{p,0}\right)^2}{8\pi} \left(\dfrac{r_p}{r_M}\right)^6 \text{,}
  \label{eq:mag_balance}
\end{equation}
where $B_{ss}$ is the stellar magnetic field extrapolated at the source surface, $R_{ss}$ is the 
radius of the source surface, $R_{orb}$ the orbital radius of the planet, and $B_{p,0}$, $r_p$, 
$r_M$ have already been defined. 

We also rewrite the pressure balance equation in the case of the Earth where the solar 
wind magnetic pressure is negligible with respect to the wind ram pressure, in this case 
we obtain:
\begin{equation}
  P_{ram,\odot}(R_{orb}^{\oplus})
  = \dfrac{\left(\dfrac{1}{2}B_{p,0}^\oplus\right)^2}{8\pi} 
\left(\dfrac{r_\oplus}{r_M^\oplus}\right)^6 \text{,}
  \label{eq:mag_balance_Earth}
\end{equation}
where $P_{ram,\odot}(R_{orb}^{\oplus})$ is the ram pressure of the solar wind at Earth, and the 
right-hand side of the equation is the same as in Eq.~(\ref{eq:mag_balance}) with Earth values.
Dividing Eq.~(\ref{eq:mag_balance}) by Eq.~(\ref{eq:mag_balance_Earth}) yields:
\begin{equation}
  \dfrac{r_M}{r_p} = \dfrac{r_M^\oplus} {r_\oplus}
  \dfrac{\left[ 8\pi P_{ram,\odot}(R_{orb}^{\oplus}) \right]^{1/6}}
  {\left(f_1 f_2 B_{\star}\right)^{1/3} \left(\dfrac{R_{ss}}{R_{orb}} \right)^{2/3}}
  \left( \dfrac{B_{p,0}}{B_{p,0}^\oplus} \right)^{1/3}
  \label{eq:rmag}
\end{equation}
In this equation, we have expressed the value of the magnetic field at the source surface as:
\begin{equation}
  B_{ss} = f_1 f_2 B_{\star}\text{,}
  \label{eq:Bss}
\end{equation}
where $f_1=B_{ZDI}/B_\star$ represents the fraction of the stellar surface magnetic 
``flux'' that can be detected through ZDI \citep{Semel89}. 
Following \citet{Reiners09} and \citet{Morin10} we take $f_1=0.2$ if the large-scale 
component of the field is dipole-dominated and $f_1=0.06$ in the multipolar case. 
The factor $f_2=B_{ss}/B_{ZDI}$ represents the ratio between the average magnetic field 
measured by ZDI and the average magnetic field at the source surface. From the sample 
of M dwarfs analyzed in V13 we find an average value of $f_2=1/15$ with little scatter 
for most stars, while a few stars with very non-axisymmetric fields have much lower 
values. This can be considered as a favourable case allowing the development of a large 
planetary magnetosphere, in the present study we consider $f_2=1/50$ as a representative 
value for this configuration. In Table~\ref{tab:rmag}, to encompass a wide range of 
physically meaningful values, we consider $f_1=0.2$, $f_2=1/15$ for a dipole-dominated 
stellar magnetic field and $f_1=0.06$, $f_2=1/50$ for a multipolar field.

By replacing numerical constants and allowing to express $R_{ss}$ and $R_{orb}$ in 
convenient units, one finally obtains the expression presented in 
Eq.~(\ref{eq:rmag_numeric}), where:
\begin{equation}
  K = \dfrac{r_M^\oplus}{r_\oplus}
      \dfrac{\left[ 8\pi P_{ram,\odot}(R_{orb}^{\oplus}) \right]^{1/6}}
            {\left( {B_{p,0}^\oplus} \right)^{1/3} \left(\dfrac{R_{ss}}{R_\star}\right)^{2/3}}
      \left(\dfrac{[1~{\rm au}]}{R_\odot}\right)^{2/3}
  \label{eq:Krmag}
\end{equation}
The value of the non-dimensional constant $K=15.48$ mentioned in the main text can be 
estimated by using the following values taken from V13: $\dfrac{r_M^\oplus}{r_\oplus}=11.7$; 
$P_{ram,\odot}(R_{orb}^{\oplus})=3.9\times10^{-9}~{\rm dyn~cm^{-}}$; 
${B_{p,0}^\oplus}=1~{\rm G}$ and $\dfrac{R_{ss}}{R_\star}=2.5$.

\subsection{Case with wind ram pressure}

In this case, Eq.~(\ref{eq:mag_balance}) becomes:
\begin{equation}
  P_{ram, {\rm Proxima}}(R_{orb})
  + \dfrac{B_{ss}^2}{8\pi} \left(\dfrac{R_{ss}}{R_{orb}}\right)^4
  = \dfrac{\left(\dfrac{1}{2}B_{p,0}\right)^2}{8\pi} \left(\dfrac{r_p}{r_M}\right)^6 \text{,}
  \label{eq:mag_balance_full}
\end{equation}
where $P_{ram, {\rm Proxima}}(R_{orb})$ is Proxima's wind ram pressure at the orbit of
Proxima~b. The expression of the magnetopause radius then becomes:
\begin{equation}
  \dfrac{r_M}{r_p} = \dfrac{r_M^\oplus} {r_\oplus}
  \left[ \dfrac{1}
    {f_w
    + \dfrac{\left(f_1 f_2 B_{\star}\right)^{2} \left(\dfrac{R_{ss}}{R_{orb}} \right)^{4}}
      {8\pi P_{ram,\odot}(R_{orb}^{\oplus})}}
  \right]^{1/6}
  \left( \dfrac{B_{p,0}}{B_{p,0}^\oplus} \right)^{1/3} \text{,}
  \label{eq:rmag_full}
\end{equation}
where the ratio $f_w = \dfrac{P_{ram, {\rm Proxima}}(R_{orb})}{P_{ram,\odot}(R_{orb}^{\oplus})}$ 
lies in the range 4--80 according to Sect.~\ref{sub:wind}.

\end{document}